\documentclass[twocolumn,preprintnumbers,superscriptaddress,amsmath,amssymb,prb]{revtex4-1}

\usepackage{graphicx}
\usepackage{dcolumn}
\usepackage{subfigure}
\usepackage{bm}
\usepackage[utf8]{inputenc}
\usepackage{comment}
\usepackage{units}
\usepackage[unicode=true, bookmarks=false, breaklinks=false, pdfborder={0 0 1},colorlinks=false]{hyperref}
\usepackage{breakurl}
\usepackage{url}
\usepackage{natbib}
\usepackage{multirow}
\usepackage{float}

\usepackage[cmyk,dvipsnames]{xcolor}
\definecolor{pacificb}{HTML}{1CA9C9}

\begin{document}

\title{Tuning skyrmions in B20 compounds by 4\textit{d} and 5\textit{d} doping}

\newcommand{\KTH}{Department of Applied Physics, School of Engineering Sciences, KTH Royal Institute of Technology, 
AlbaNova University Center, SE-10691 Stockholm, Sweden}
\newcommand{\SeRC}{SeRC (Swedish e-Science Research Center), KTH Royal Institute of Technology, SE-10044 Stockholm, Sweden}
\newcommand{\Uppsala}{Department of Physics and Astronomy, Uppsala University, Box 516, SE-75120 Uppsala, Sweden}
\newcommand{\Orebro}{School of Science and Technology, \"Orebro University, SE-701 82, \"Orebro, Sweden}
\newcommand{\Stockholm}{Department of Materials and Environmental Chemistry, Stockholm University, SE-10691 Stockholm, Sweden}
\newcommand{\UppsalaChem}{Department of Chemistry, Uppsala University, Box 538, Uppsala, SE-751 21, Sweden}
\newcommand{\PDC}{PDC Center for High Performance Computing, KTH Royal Institute of Technology, SE-100 44 Stockholm, Sweden}

\author{Vladislav Borisov}
    \affiliation{\Uppsala}
    \email[Corresponding author:\ ]{vladislav.borisov@physics.uu.se}

\author{Qichen Xu}
    \affiliation{\KTH}
    \affiliation{\SeRC}

\author{Nikolaos Ntallis}
    \affiliation{\Uppsala}
    
\author{Rebecca Clulow}
    \affiliation{\UppsalaChem}

\author{Vitalii Shtender}
    \affiliation{\UppsalaChem}    

\author{Johan Cedervall}
    \affiliation{\Stockholm}

\author{Martin Sahlberg}
    \affiliation{\UppsalaChem}

\author{Kjartan Thor Wikfeldt}
    \affiliation{\PDC}

\author{Danny Thonig}
    \affiliation{\Orebro}
    \affiliation{\Uppsala}

\author{Manuel Pereiro}
    \affiliation{\Uppsala}

\author{Anders Bergman}
    \affiliation{\Uppsala}

\author{Anna Delin}
    \affiliation{\KTH}
    \affiliation{\SeRC}

\author{Olle Eriksson}
    \affiliation{\Uppsala}
    \affiliation{\Orebro}
    
\date{\today}

\begin{abstract}
Skyrmion stabilization in novel magnetic systems with the B20 crystal structure is reported here, primarily based on theoretical results. The focus is on the effect of alloying on the 3\textit{d} sublattice of the B20 structure by substitution of heavier 4\textit{d} and 5\textit{d} elements, with the ambition to tune the spin-orbit coupling and its influence on magnetic interactions. State-of-the-art methods based on density functional theory are used to calculate both isotropic and anisotropic exchange interactions. Significant enhancement of the Dzyaloshinskii-Moriya interaction is reported for 5\textit{d}-doped FeSi and CoSi, accompanied by a large modification of the spin stiffness and spiralization.
Micromagnetic simulations coupled to atomistic spin-dynamics and \textit{ab initio} magnetic interactions reveal a helical ground state and field-induced skyrmions for all these systems. Especially small skyrmions $\sim\unit[50]{nm}$ are predicted for Co$_{0.75}$Os$_{0.25}$Si, compared to $\sim\unit[148]{nm}$ for Fe$_{0.75}$Co$_{0.25}$Si. Convex-hull analysis suggests that all B20 compounds considered here are structurally stable at elevated temperatures and should be possible to synthesize. This prediction is confirmed experimentally by synthesis and structural analysis of the Ru-doped CoSi systems discussed here, both in powder and in single-crystal forms.
\end{abstract}

\maketitle

\section{Introduction}

\begin{figure}
\centering
\includegraphics[width=0.9\columnwidth]{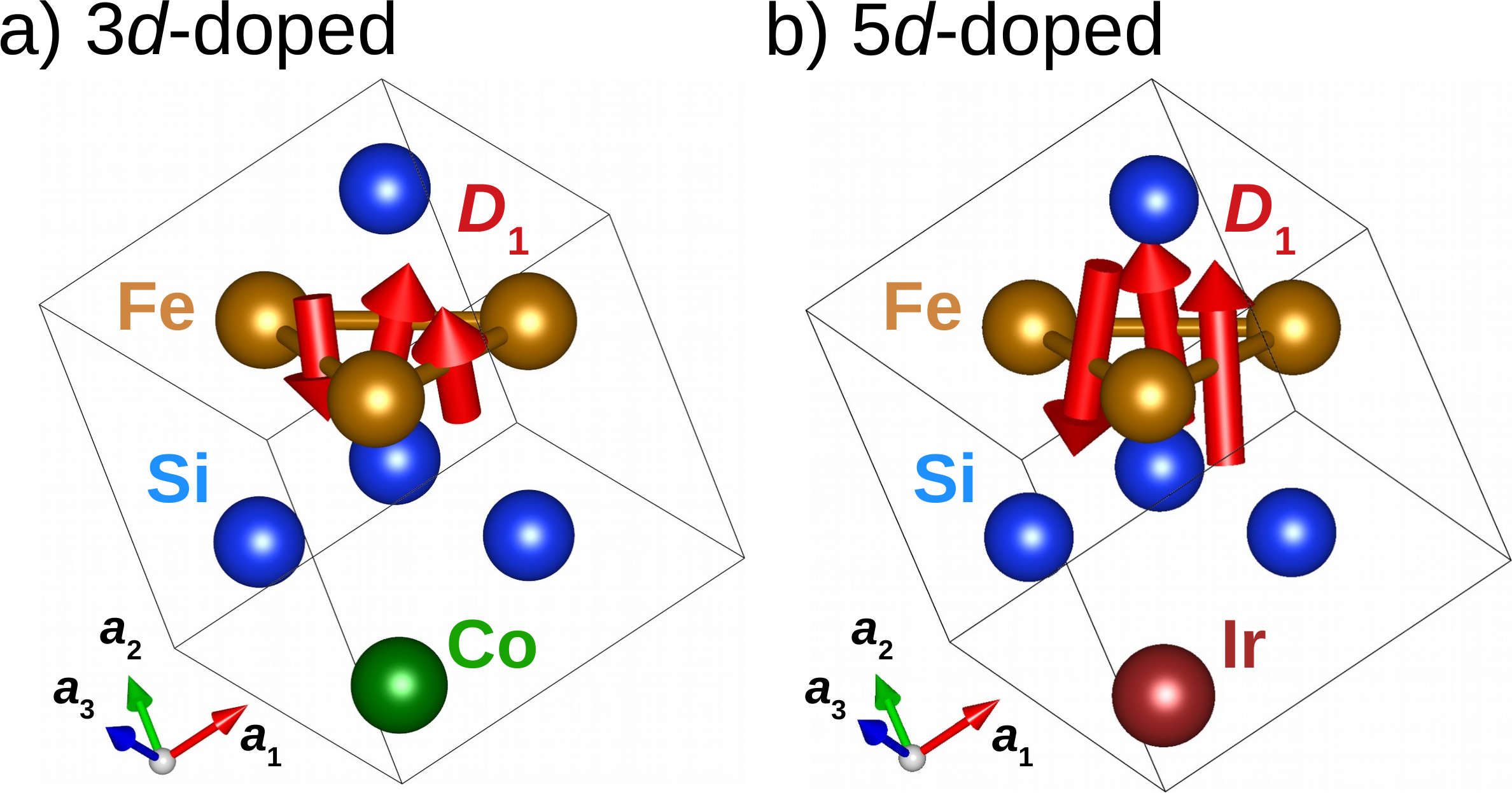}
\caption{Dzyaloshinskii-Moriya vectors $(\vec{D}_1)$ for the nearest-neighbor Fe-Fe bonds in the doped B20 compounds: a) Fe$_{0.75}$Co$_{0.25}$Si and b) Fe$_{0.75}$Ir$_{0.25}$Si. The direction and size of the DM vectors are given by red arrows, where the arrow length scales with the DMI strength.
}
\label{f:D1_vectors}
\end{figure}

Alloy mixtures can sometimes show unexpected properties compared to those of pure components, that are used to make up the alloyed state. This is, for example, the case for the Kondo insulator FeSi \cite{Aeppli1992} and the diamagnetic metal CoSi \cite{Wernick1972}, both having the chiral crystal structure of the B20 type (Figure~\ref{f:D1_vectors}). This cubic structure belongs to the $P2_13$ space group and its chirality becomes apparent when the nearest-neighbor Si atoms for each magnetic transition metal (\textit{TM})-site are considered in terms of the direction of \textit{TM}-Si bonds. While pure FeSi and CoSi do not show any long-range magnetism, an alloyed mixture of them, Fe$_{1-x}$Co$_x$Si, reveals surprisingly a helical magnetic order in a wide range of concentrations $x$ \cite{Manyala2000,Onose2005}. The dependence of these two quantities as well as the helical spatial period on the Co concentration is rather non-trivial, with the maximal Curie temperature of $\unit[50]{K}$ and an ordered moment $\sim\unit[0.2]{\mu_\mathrm{B}/\mathrm{f.u.}}$ around $x=\unit[40]{\%}$. For a certain range of concentrations, the application of an external magnetic field can induce a skyrmion lattice in these alloys \cite{Muenzer2010,Yu2010}, similarly to the B20 compound MnSi \cite{Muehlbauer2009}.

The skyrmionic properties of such systems rely on the interplay between the Heisenberg and Dzyaloshinskii-Moriya \cite{Dzyaloshinsky1958,Moriya1960} (DM) exchange interactions. The DM interaction (DMI) can actually be significant even if the alloy does not contain heavy elements, which usually contribute to the DMI via their large spin-orbit coupling. However, it is natural to expect that the DMI would be further enhanced, if the system could contain 4\textit{d} or 5\textit{d} elements, since heavier elements are known to have larger spin-orbit interaction. This idea is corroborated by the experimental observation of robust skyrmions in transition metal multilayers with heavy elements, e.g., Fe/Ir(111) \cite{Heinze2011}, Pt/Co/Ta \cite{Wang2019} and Ir/Fe/Co/Pt \cite{Soumyanarayanan2017}. Several attempts have been undertaken to synthesize and study the magnetic properties of B20 compounds doped by heavy elements, for example, Rh-doped MnGe \cite{Sidorov2018} as well as Ir-doped MnSi \cite{Dhital2017} and FeSi \cite{Sales1994}. From the study of Mn$_{1-x}$Ir$_x$Si \cite{Dhital2017}, it was, however, unclear whether there is any significant change of the DMI with Ir doping and more accurate ways of extracting the DMI strength from experiment were claimed to be necessary. Concerning Fe$_{1-x}$Ir$_x$Si \cite{Sales1994}, its magnetic properties have been studied experimentally only up to $x=0.1$ where no long-range order was observed. In general, the effect of 4\textit{d} and 5\textit{d} doping on the magnetic interactions and skyrmions in different B20 compounds has not been systematically explored so far.

In this work, we study theoretically the possibility of improving the skyrmionic properties of FeSi- and CoSi-based B20 compounds by means of 4\textit{d} and 5\textit{d} doping, compared to 3\textit{d} doping. The investigation uses \textit{ab initio} electronic structure theory with a focus on magnetic moments, isotropic Heisenberg exchange as well as anisotropic interactions (symmetric exchange and DMI). We find a significant enhancement of the calculated DM interaction in FeSi, when 25\% Ir is alloyed on the Fe sublattice (Figure~\ref{f:D1_vectors}), and in Ru- and Os-doped CoSi. This causes the ratio between the DM and Heisenberg interactions to increase substantially, especially for CoSi-based systems. Using these \textit{ab initio} interactions, our simulations of the magnetization dynamics show a helical magnetic ground state in zero field and the formation of magnetic skyrmions with topological number $1$ when a magnetic field is applied. Convex-hull analysis shows that the considered doped B20 compounds are structurally stable and should be possible to synthesize at elevated temperature, around $\unit[1000]{K}$, where the mixing entropy increases the stability. This result is corroborated by a successful synthesis of Co$_{1-x}$Ru$_{x}$Si single crystals that we achieved, which is a new class of B20 systems not reported before in literature.

\section{Results}

\newcommand{\FS}{Fe$_{0.75}$\textit{TM}$_{0.25}$Si}
\newcommand{\CS}{Co$_{0.75}$\textit{TM}$_{0.25}$Si}
\newcommand{\FSx}{Fe$_{1-x}$\textit{TM}$_{x}$Si}
\newcommand{\CSx}{Co$_{1-x}$\textit{TM}$_{x}$Si}

\subsection{Heisenberg and DM interactions}

The effect of 3\textit{d}, 4\textit{d} and 5\textit{d} alloying on the magnetic interactions in FeSi- and CoSi-based B20 compounds is considered first. The main result is a significant enhancement of the DM interaction, which is most pronounced for the 5\textit{d} doping (a comparison between Co- and Ir-doped FeSi is illustrated in Figure~\ref{f:D1_vectors}, where the size of the interaction is represented by the length of the arrow). Below, we elaborate on this result and discuss further aspects of the exchange interactions of doped B20 compounds.

\begin{figure}
\centering
\includegraphics[width=0.99\columnwidth]{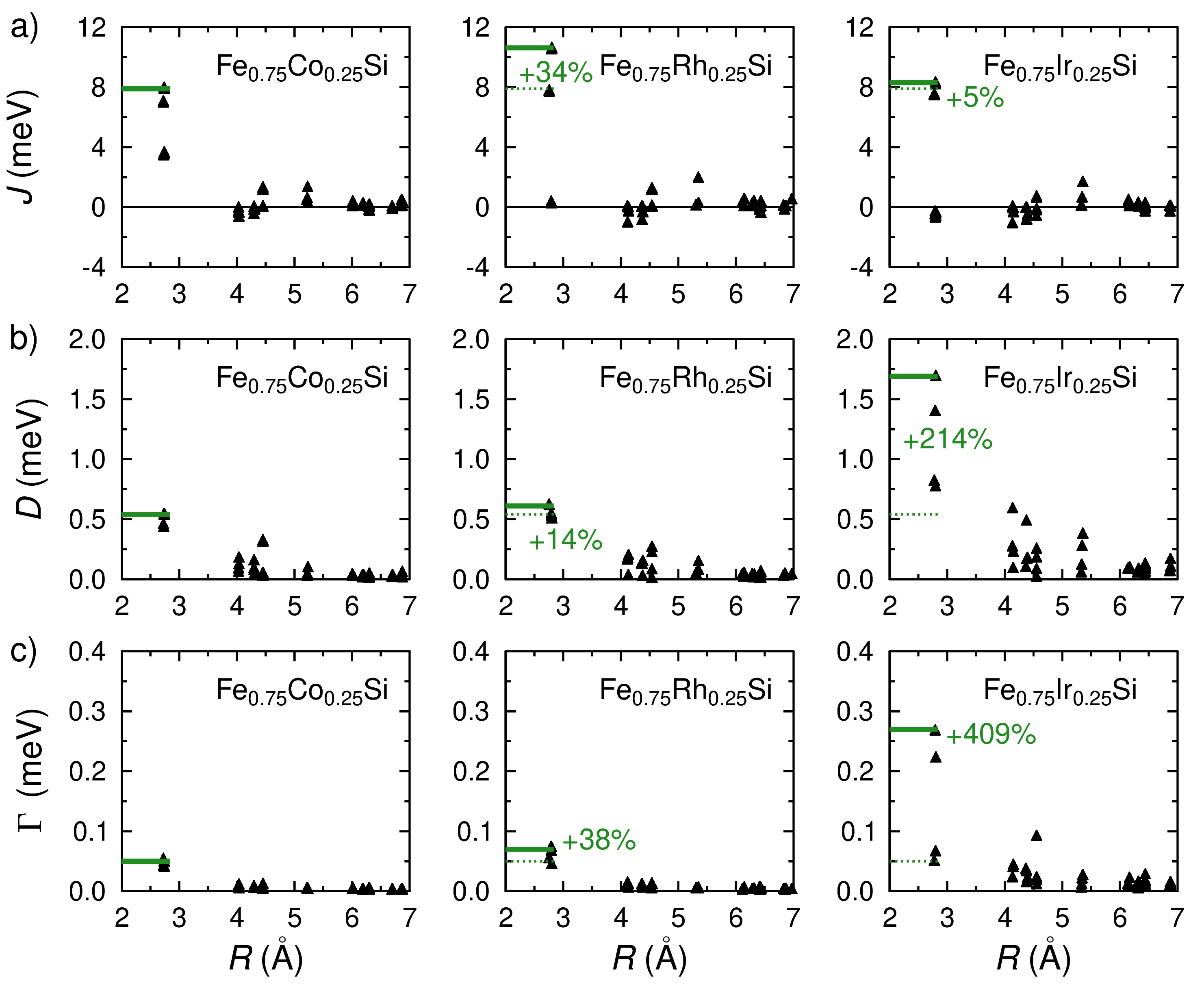}
\caption{Evolution of a) Heisenberg, b) Dzyaloshinskii-Moriya and c) symmetric anisotropic interactions in the doped series Fe$_{0.75}$\textit{TM}$_{0.25}$Si (\textit{TM}\,=\,Co, Rh, Ir). The $D$ and $\Gamma$ parameters are defined as $D = \sqrt{D_x^2 + D_y^2 + D_z^2}$ and $\Gamma=\sqrt{\Gamma_{xy}^2 + \Gamma_{xz}^2 + \Gamma_{yz}^2}$. The magnetic interactions are calculated for the ferromagnetic reference configuration and are plotted as functions of the distance between the interacting spins. The horizontal green lines mark the maximal value of interaction for each case, and the relative change for the 4$d$- and 5$d$-doped cases compared to the 3$d$-doped case is given in percent.}
\label{f:Jij_FeTMSi}
\end{figure}

\begin{figure}
\begin{centering}
\includegraphics[width=0.99\columnwidth]{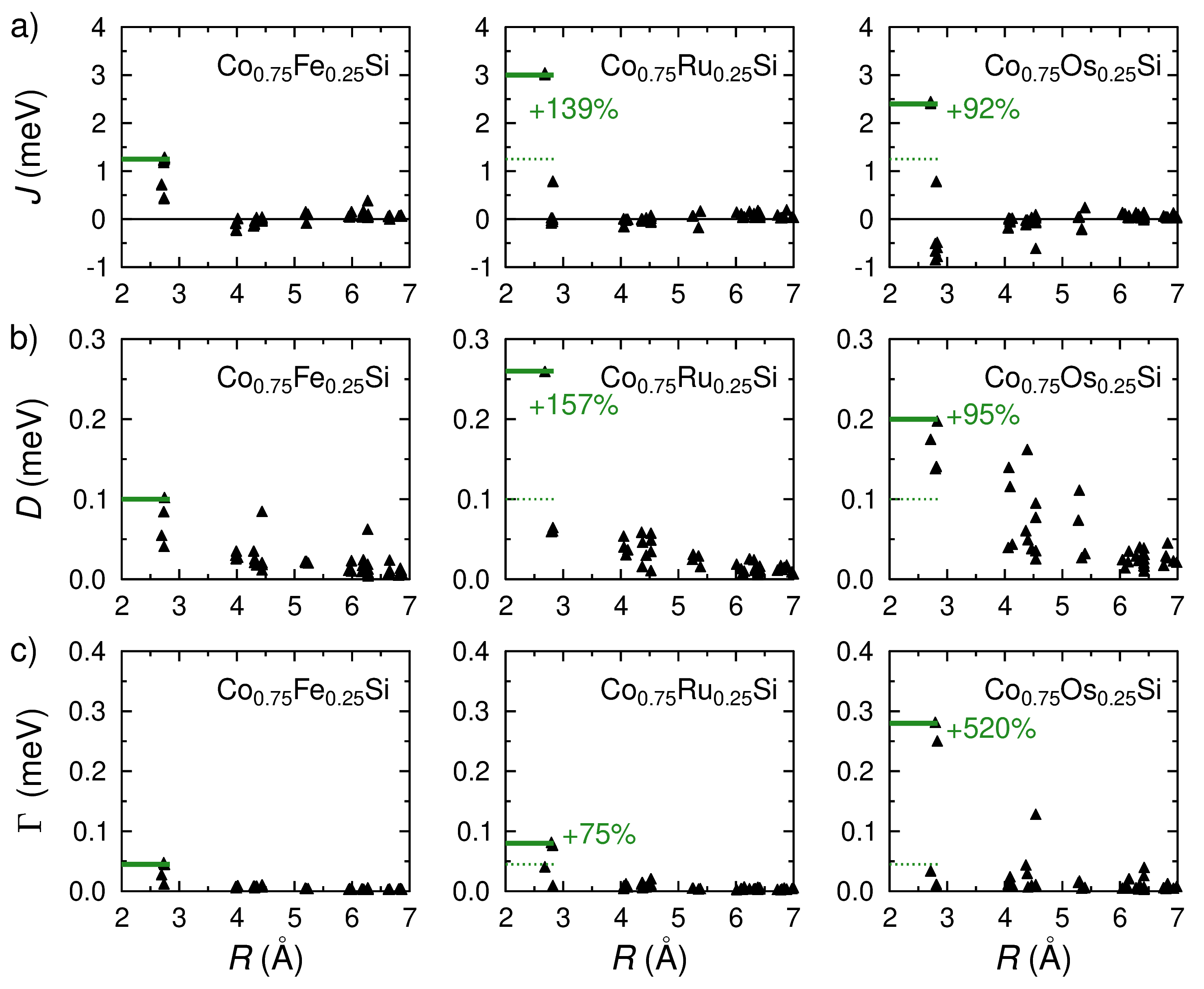}
\end{centering}
\caption{Evolution of a) Heisenberg, b) Dzyaloshinskii-Moriya and c) symmetric anisotropic  interactions in the doped series Co$_{0.75}$\textit{TM}$_{0.25}$Si (\textit{TM}\,=\,Fe, Ru, Os). The $D$ and $\Gamma$ parameters are defined as $D = \sqrt{D_x^2 + D_y^2 + D_z^2}$ and $\Gamma=\sqrt{\Gamma_{xy}^2 + \Gamma_{xz}^2 + \Gamma_{yz}^2}$. The magnetic interactions are calculated for the ferromagnetic reference configuration and are plotted as functions of the distance between the interacting spins. The horizontal green lines mark the maximal value of interaction for each case, and the relative change for the 4$d$- and 5$d$-doped cases compared to the 3$d$-doped case is given in percent.}
\label{f:Jij_CoTMSi}
\end{figure}
\begin{figure}
\centering
\includegraphics[width=0.9\columnwidth]{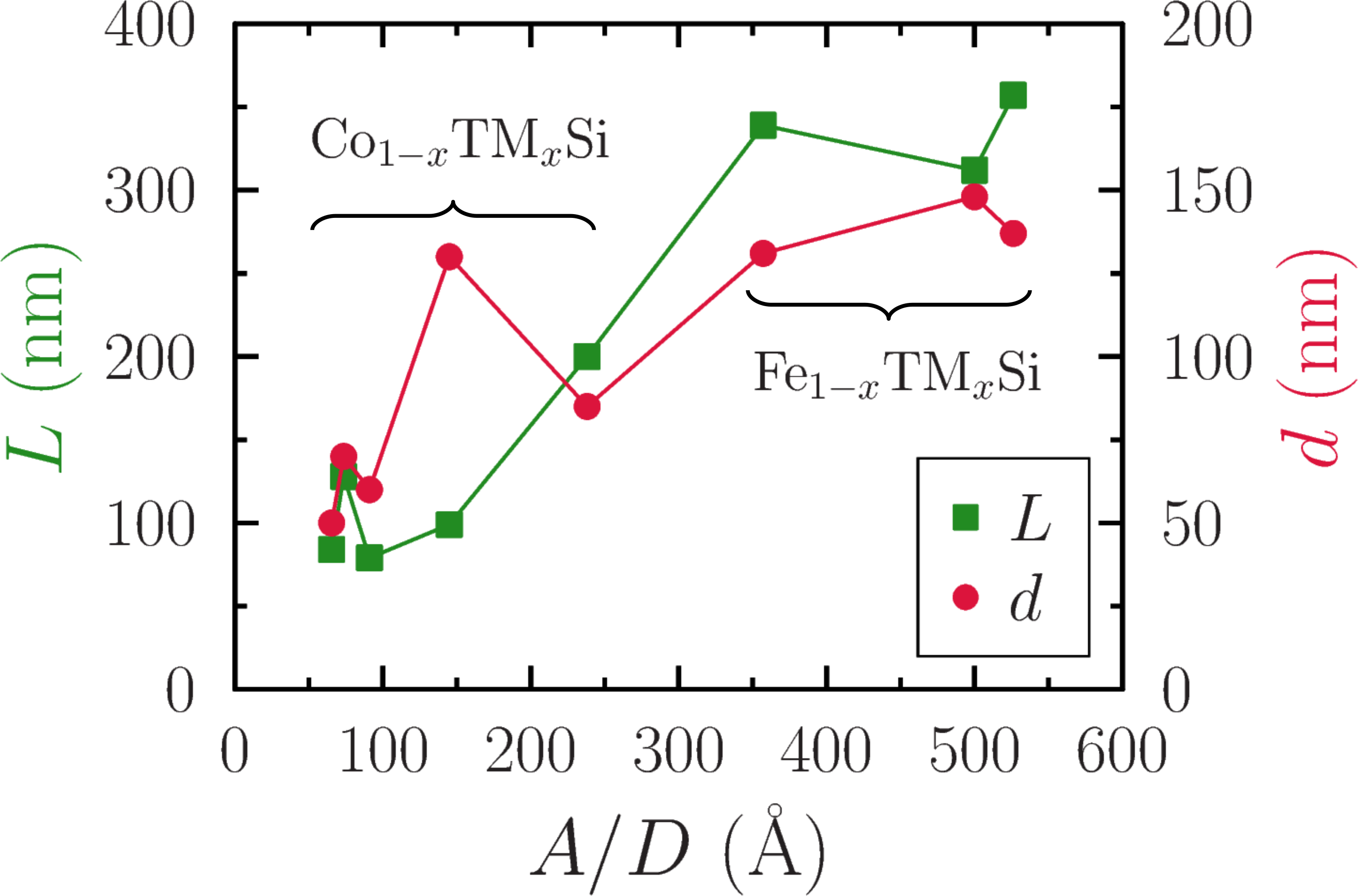}
\caption{Relation between the spiral wave-length (left axis, green), skyrmion size (right axis, red) and the $\nicefrac{A}{D}$ ratio for the doped B20 compounds.}
\label{fig:AtoD_ratios}
\end{figure}

The calculated element-specific magnetic moments, Heisenberg exchange $(J)$, DM interactions ($D$) and anisotropic symmetric exchange ($\Gamma^{\alpha\beta}$) are shown in Figure~\ref{f:Jij_FeTMSi} for 25\%-doping of Co, Rh and Ir in FeSi (similar Figure~\ref{f:Jij_CoTMSi} shows doped CoSi). Significant interactions are observed for spins within a short distance not extending much beyond $\unit[6]{\AA}$. The short-range character of the interactions is somewhat surprising, given the metallic nature of these systems. On the other hand, for some compounds the simplistic picture with only nearest-neighbor interaction would be not sufficient, since a few further neighboring shells show non-negligible interactions. In terms of the magnitude of the Heisenberg exchange, the Co- and Ir-doped cases are similar (Figure~\ref{f:Jij_FeTMSi}a), while Rh doping increases the maximal value of the Heisenberg exchange by $\sim\!\unit[34]{\%}$ (Fig.~\ref{f:Jij_FeTMSi}a). This can be explained by the relatively large Fe moment of the Rh doped system ($\unit[0.97]{\mu_\mathrm{B}}$ in Table~1 in SI), which is larger than the Co- and Ir-doped systems. Note that in our formalism the length of the magnetic moments are incorporated in the exchange interaction, see Equation~(\ref{e:general_Heisenberg_model}). In contrast, the DM interaction is only slightly affected by Rh doping compared to Co doping and a significant enhancement of the largest $D_{ij}$ by $\sim\!\unit[214]{\%}$ is only achieved in the Ir-doped case (Figure~\ref{f:Jij_FeTMSi}b). The micromagnetic $\nicefrac{D}{A}$ ratio (Table~1 in SI and the inverse ratio $\nicefrac{A}{D}$ in Figure~\ref{fig:AtoD_ratios}) characterizes the relative energy scales of the DM and Heisenberg interactions and is seen to increase in the series 3\textit{d}-4\textit{d}-5\textit{d}, similar to the nearest-neighbor ratio of atomic interactions $\nicefrac{D_1}{J_1}$ (Table~1). Such an enhancement of the DM interaction can lead to skyrmions with smaller geometry. Interestingly, the symmetric anisotropic exchange is relatively small in 3\textit{d}- and 4\textit{d}-doped FeSi but shows an unexpected increase by 409\% for the 5\textit{d}-doped system (Figure~\ref{f:Jij_FeTMSi}c). However, this exchange interaction is still below $\unit[0.3]{meV}$ and is not expected to compete with the DM interaction.

In the CoSi-based B20 compounds (Figure~\ref{f:Jij_CoTMSi}a), 25\%-doping of Fe, Ru and Os results in a Heisenberg exchange that is factor of three or four smaller than in the FeSi-based systems (Figure~\ref{f:Jij_FeTMSi}a). This is related to the significantly weaker magnetic moments of the cobalt-rich systems, as shown in Table~1. We also observe that the Heisenberg exchange $J$ in Co$_{0.75}$\textit{TM}$_{0.25}$Si is much more sensitive to doping than in Fe$_{0.75}$\textit{TM}$_{0.25}$Si. Compared to the Fe-doped CoSi, the nearest-neighbor (NN) $J$-parameter is enhanced by 139\% in the Ru-doped CoSi and by 92\% in the Os-doped case. Surprisingly, also the DM interaction for some of the NN bonds is increased by up to 157\% due to the 4\textit{d} dopants, while the 5\textit{d} doping provides an increase of 95\% but for several interactions within $\unit[5.5]{\AA}$ (Figure~\ref{f:Jij_CoTMSi}b). It is worth noticing that not only the magnitude but also the direction of the DM vectors is important for non-collinear magnetic textures and this fact is taken into account in micromagnetic expression (\ref{e:micromagnetic_parameters}). The $\nicefrac{D}{A}$ ratio increases considerably for the CoSi systems, reaching $0.015$ for the Os-doped case, which together with the results for FeSi systems suggests that 5\textit{d} doping is especially effective for tuning the chiral magnetic interactions in B20 compounds. This conclusion holds for the symmetric anisotropic exchange $\Gamma^{\alpha\beta}$ as well, described by Equation~(\ref{e:symmetric_exchange}). The magnitude of this symmetric exchange increases by 520\% for the Os-doped case compared to the Fe-doped system, remaining, however, below $\unit[0.3]{meV}$, similarly to the FeSi systems. The difference is that the DM interactions ($D$) and symmetric exchange ($\Gamma$) are on the same scale for Co$_{0.75}$Os$_{0.25}$Si (Figure~\ref{f:Jij_CoTMSi}b,c) which may, in principle, produce interesting magnetic effects. The comparable magnitudes of $D$ and $\Gamma$ make Co$_{0.75}$Os$_{0.25}$Si a unique system, since the symmetric exchange $\Gamma$ is rarely discussed in the literature, in contrast to the DM interaction.

The magnetic ordering temperatures were calculated from Monte Carlo simulations and Binder cumulant analysis (see Figure~S\ref{fig:magnetization} and S\ref{fig:cumulants}), and are summarized in Table~1 in SI. We see that the ordering temperature $T_c$ is slightly increased by 4\textit{d} doping both for FeSi and for CoSi compounds, while the 5\textit{d} doping reduces $T_c$ moderately due to smaller induced moments on 5\textit{d} atoms and consequently weaker magnetic exchange. While the $T_c$ for CoSi-based systems is around $\unit[52-56]{K}$, the critical temperature for FeSi-based systems is overestimated in the DFT calculations, as we discuss in more detail in section~3 of SI. Based on the previous experimental studies of Fe$_{1-x}$Co$_x$Si one may expect the actual $T_c$ that would be measured for the studied B20 compounds in the future to be below $\unit[30]{K}$. For completeness we have also calculated the adiabatic magnon spectra of all investigated systems (Figure~\ref{fig:ams}), with the most distinct result being a significant magnon band gap around $\unit[1.5]{meV}$ for Fe$_{0.75}$Ir$_{0.25}$Si, that may have implications for magnonics.

\subsection{Micromagnetic simulations}

\begin{figure}
\centering
\includegraphics[width=0.9\columnwidth]{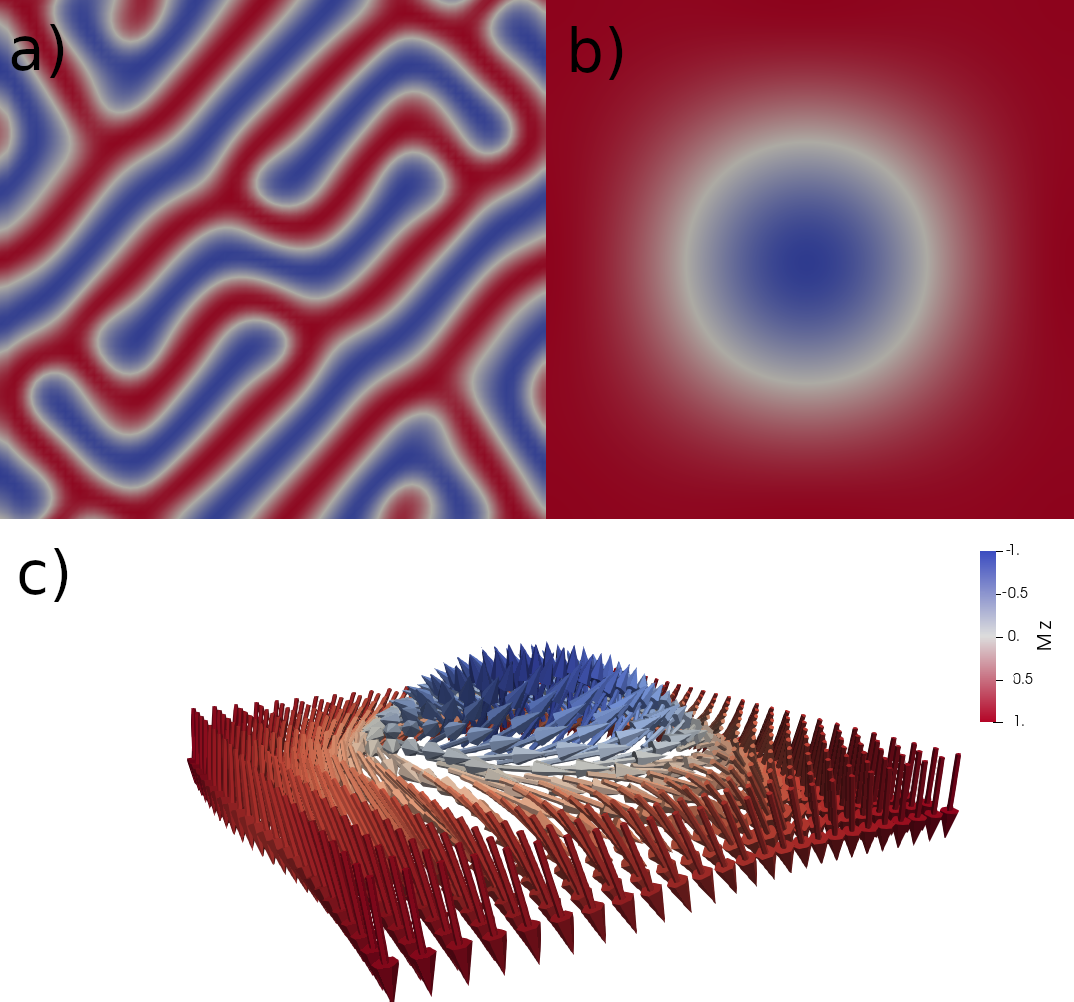}
\caption{a) Helical magnetic state of Fe$_{0.75}$Ir$_{0.25}$Si and b,c) stabilized skyrmion after the application of magnetic field of $\unit[5]{mT}$. b) Real-space slice of the magnetization showing one skyrmion. c) The calculated magnetization distribution in the skyrmion shown by arrows. The color indicates the $z$-component of the magnetic moment, such that the blue (red) color represents the moments pointing fully along the $+z$-direction ($-z$-direction). The total size of the simulation box is a) $(3\times3\times1)\,\mu\mathrm{m}$ and b) $(0.4\times0.4\times0.1)\,\mu\mathrm{m}$.}
\label{fig:micromagnetic}
\end{figure}
Since the wavelength of the spin-spiral ground state of known B20 compounds is of the order of tens or hundreds of nanometers and is too large for atomistic spin dynamics simulations, we have performed micromagnetic simulations with the multiscale module $\mu$-\textsc{ASD} \cite{pereiro} of the \textsc{UppASD} code \cite{uppasd,Eriksson2017}. The magnetic ground states of the {\FS} and {\CS} compounds in zero magnetic field are depicted in Figure~\ref{fig:micromagnetic} and~\ref{fig:magnetic_textures} (in SI) in $(3\times 3\times 1)\,\mu\mathrm{m}$ region, where the helical domain structure can be appreciated. It is apparent that Rh-doped FeSi (Figure~\ref{fig:magnetic_textures}b) has the largest spiral wave-length ($L = \unit[357]{nm}$) while Fe$_{0.75}$Co$_{0.25}$Si and Fe$_{0.75}$Ir$_{0.25}$Si reveal similar spiral patterns (Table~\ref{tab:micro_param} in SI). We find that the calculated magnetic wave-length $L$ is roughly proportional to the $\nicefrac{A}{D}$ ratio, as illustrated by Figure~\ref{fig:AtoD_ratios}. We notice that $D/A$ is visibly increased for the Ir-doped FeSi compared to the other two systems (see Table~1), indicating a stronger relative contribution of the DM interaction to the magnetic properties. For doped CoSi, the length scale of the calculated helical ground states in Figure~\ref{fig:magnetic_textures} is substantially shorter, which is a consequence of larger $D/A$ ratios. Interestingly, the trends shown by the micromagnetic $|D/A|$ and nearest-neighbor atomistic $D_1/J_1$ ratios are different which indicates that an accurate picture of magnetic phenomena must take into account not just the nearest but also further neighbor shells and the direction of the DM vectors that affects $\vec{D}_{ij}\cdot\vec{R}_{ij}$ in Equation~(\ref{e:micromagnetic_parameters}).
    
As an example of a usual procedure to stabilize skyrmions in chiral materials, we show in Figure~\ref{fig:micromagnetic} the skyrmion state found in Fe$_{0.75}$Ir$_{0.25}$Si after the application of an external field of $\unit[5]{mT}$. The system does not show the formation of a skyrmion lattice but rather of individual Bloch type skyrmions, which could be, in principle, manipulated by means of spin currents. The diameter of the stabilized skyrmions is around $\unit[131]{nm}$, while the smallest skyrmions are predicted for Co$_{0.75}$Os$_{0.25}$Si. As expected, we find an increasing linear trend of the skyrmion size as a function of the $\nicefrac{A}{D}$ ratio (Figure~\ref{fig:AtoD_ratios}). The skyrmion number is $1$ for all compounds studied in this work and, interestingly, the formation of skyrmions in the Co-based B20 systems requires higher magnetic fields compared to FeSi doped compounds.

\subsection{Phase- and structural stability from theory}

So far, we have discussed the magnetic properties of doped B20 compounds predicted by DFT calculations combined with effective spin-Hamiltonians and magnetisation dynamics. Since most of these systems have not been studied experimentally yet, it is important to estimate their structural stability, which would indicate whether it is possible to synthesize these doped systems. The previously studied Fe$_{1-x}$Co$_{x}$Si system, that is known from experiments to be stable for $0\leq x \leq 1$ \cite{Onose2005}, can be a reference for validating the chosen theoretical approach and to assess its accuracy.

\begin{figure}
\centering
\includegraphics[width=0.99\columnwidth]{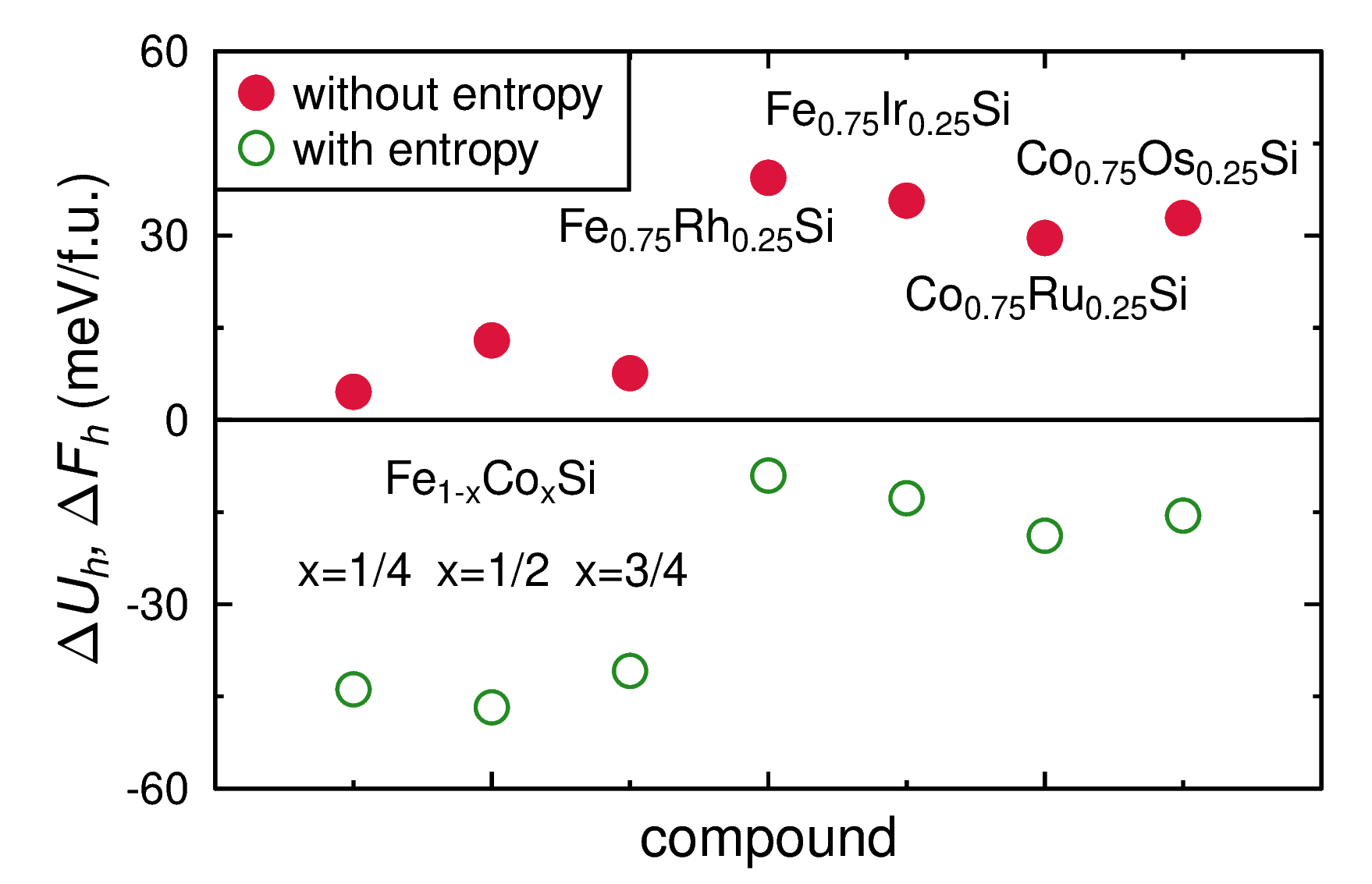}
\caption{Distance from the convex hull for doped B20 compounds (from left to right): Fe$_{1-x}$Co$_{x}$Si ($x=\nicefrac{1}{4},\nicefrac{1}{2},\nicefrac{3}{4}$), Fe$_{0.75}$Rh$_{0.25}$Si, Fe$_{0.75}$Ir$_{0.25}$Si, Co$_{0.75}$Ru$_{0.25}$Si and Co$_{0.75}$Os$_{0.25}$Si. Filled and open symbols correspond to zero-temperature distance $\Delta U_h$ (without entropy) and finite-temperature $\Delta F_h$ (with mixing entropy, $T=\unit[1000]{K}$). $\Delta F_h < 0$ indicates a phase- and structural stability.}
\label{f:distance_from_hull}
\end{figure}

For all studied B20 compounds, we obtain a negative formation energy around $\unit[-0.5]{eV/f.u.}$ (Figure~\ref{f:formation_energy}) which indicates a stability with respect to the decomposition into the pure elements. This condition is necessary but not sufficient, since it does not exclude the decomposition into more complex competing phases, for example, FeSi, CoSi, Fe$_3$Si, CoSi$_2$ etc.~in case of Fe$_{1-x}$Co$_{x}$Si (further phases in Figure~\ref{f:convex_hull_diagrams}e). A large number of possible decomposition scenarios can be systematically taken into account by means of a convex hull analysis, the results of which we present in Figure~\ref{f:distance_from_hull} and \ref{f:convex_hull_diagrams}.

In Figure~\ref{f:convex_hull_diagrams}, the corners of the triangles correspond to the pure constituent elements. Any point inside the triangle is characterized by the projections onto the triangle edges which indicate the amount of each element in the compound. The associated formation energy $\Delta U$ or free energy $\Delta F$ are either above or below the convex hull. The distance from the hull ($\Delta U_h$ or $\Delta F_h$) is the main parameter which allows to judge how stable a particular compound is. Our results summarized in Figure~\ref{f:distance_from_hull} suggest that $\Delta F_h$ are negative for the Fe$_{1-x}$Co$_{x}$Si compounds with $x = \nicefrac{1}{4}$, $\nicefrac{1}{2}$ and $\nicefrac{3}{4}$ which are known to be stable. Since these systems are known to exist, this part of the calculation serves to validate the chosen approach for studying phase- and structural stability. It should be noted that $\Delta U_h > 0$ suggests an instability of the system under study. However, the mixing entropy contribution, $\Delta S$, leads to a negative $\Delta F_h$ at the chosen temperature, $T=\unit[1000]{K}$, implying that high-temperature synthesis should be possible. This is indeed the case, as we elaborate in the next paragraph. The mixing entropy, in general, improves the stability of doped systems where a random distribution of impurities is assumed, because $\Delta S$ is always positive. Similarly, the 4\textit{d}- and 5\textit{d}-doped systems are expected to be stable ($\Delta F_h < 0$, see Figure~\ref{f:distance_from_hull}) when the mixing entropy contribution at $T=\unit[1000]{K}$ is taken into account.

\subsection{Synthesis and structure analysis}

A series of samples with composition Co$_{1-x}$Ru$_x$Si and $x$ = 0.25, 0.5 and 0.75 was synthesised by arc melting and analysed by X-ray diffraction. The powder X-ray diffraction patterns (Figure~\ref{fig:Co0.75Ru0.25Si_Rietveld_Plot}-\ref{fig:Co0.25Ru0.75Si_Rietveld_Plot}) confirms that the new compositions  adopt the cubic B20 structure type ($P$2\textsubscript{1}3 space group). The diffraction patterns did not show any evidence of a secondary phase in Co\textsubscript{0.75}Ru\textsubscript{0.25}Si and Co\textsubscript{0.5}Ru\textsubscript{0.5}Si samples whilst the Co\textsubscript{0.25}Ru\textsubscript{0.75}Si contained around 2 wt\% of Ru\textsubscript{2}Si\textsubscript{3} as previously reported in undoped RuSi \cite{Perring1999}. Crystal structure refinement from the powder X-ray diffraction (PXRD) data resulted in unit cell parameters $a$ = 4.5270 (1) \AA{}, 4.5919 (1) \AA{} and 4.6528 (1) \AA{} for compositions with $x$ = 0.25, 0.5 and 0.75 respectively. The unit cell parameter increases linearly with Ru content, see Figure~\ref{fig:CoRuSi Unit Cells.jpg}, indicating the inclusion of Ru into the crystal structure and the refined occupancies of the Co/Ru sites were  0.8/0.2, 0.6/04 and 0.3/0.7 respectively. The atomic coordinates and occupancies are given in full in Tables \ref{tab:Atomic_Coordinates_Co0.75}-\ref{tab:Atomic_Coordinates_Co0.25} in SI and the Rietveld refinement plots in Figure~\ref{fig:Co0.75Ru0.25Si_Rietveld_Plot}-\ref{fig:Co0.25Ru0.75Si_Rietveld_Plot}. The composition of the sample was also investigated using energy dispersive X-ray spectroscopy (EDS) analysis giving values within 2 at\% of the nominal nominal composition (Table~\ref{tab:EDS Data} in SI). Small regions of a secondary phase were observed in the Co\textsubscript{0.25}Ru\textsubscript{0.75}Si sample which corresponded to the Ru\textsubscript{2}Si\textsubscript{3} present in the X-ray diffraction pattern. The structure type and space group of Co\textsubscript{0.75}Ru\textsubscript{0.25}Si were also confirmed by the single crystal X-ray diffraction data, the refined unit cell parameter was slightly larger than from PXRD, $a = \unit[4.5629(6)]{\AA}$ and coincides with a slightly higher refined Ru content of 0.39. The full crystallographic data and refinement details are shown in Table~\ref{tab:Co0.61Ru0.39Si_singlecrystal} and \ref{tab:C00.75Ru0.25Si Single crystal Coordinates} in SI.

\section{Conclusion}

Using state-of-the-art theoretical methods we have studied the magnetic interactions and topological textures in 3\textit{d}-, 4\textit{d}- and 5\textit{d}-doped B20 compounds based on FeSi and CoSi. The most significant finding is that 5\textit{d} doping (by Ir or Os) enhances the Dzyaloshinskii-Moriya interaction considerably, especially for the CoSi systems where the smallest skyrmions around 50\,nm are predicted for Co$_{0.75}$Os$_{0.25}$Si. Interestingly, CoSi-based compounds require larger magnetic field for skyrmion formation and, in general, show smaller skyrmions compared to the FeSi systems. The magnetic ordering temperature is found to be sensitive to doping, in agreement with the literature, and is expected to be below 30\,K. Based on the convex-hull analysis, we predict that all the studied B20 compounds are structurally stable and can be synthesized at high temperature. For Co$_{1-x}$Ru$_x$Si with $x=\frac14, \frac12, \frac34$, we actually report the first successful synthesis and characterization.

Our work predicts that 4\textit{d}- and 5\textit{d}-doped B20 compounds are promising as systems holding skyrmions, with varying size of the skyrmions depending on system. Our results suggest that Co$_{0.50}$Ru$_{0.50}$Si is optimal in terms of the magnetization, $\nicefrac{D}{A}$ ratio and skyrmion diameter, compared to other studied systems. Experimental synthesis shows that several of these compounds can be formed, in particular the Co$_{1-x}$Ru$_x$Si-system with up to 75\% Ru concentration. The results reported here hence demonstrate several new compounds in the family of materials crystallizing in the B20 structure. Theory suggests that several of these are magnetic and can hold magnetic structures with non-trivial topology.

\section{Methods}
\vspace{5pt}
The essential aspects of our theoretical simulations are described further below, while all the technical details are discussed in the supporting information (SI).

The structural, electronic and magnetic properties of doped B20 compounds Fe$_{1-x}$TM$_x$Si and Co$_{1-x}$TM$_x$Si are studied theoretically using density functional theory (DFT) \cite{Hohenberg1964,Kohn1965} within the generalized-gradient approximation in the PBE parameterization \cite{PBE1996}. Previous studies, including our most recent work \cite{Borisov2021}, suggest that electronic correlations are not crucial for describing the magnetic properties of B20 compounds. Since we are mostly interested in general trends, we do not take into account additional correlation effects beyond GGA in this work.

The doping is simulated within the supercell approach where one of the four Fe or Co magnetic sites in the unit cell is replaced by another 3$d$, 4$d$ or 5$d$ element (Figure~\ref{f:D1_vectors}), which are Co, Rh, Ir for FeSi and Fe, Ru, Os for CoSi. The supercell structure is optimized using DFT, as available in the VASP code \cite{Kresse1996}, and the calculated structural parameters are summarized in Table~1 in the SI. Ferromagnetic order is imposed in these and the subsequent DFT calculations. Electronic properties, magnetic moments and interatomic exchange interactions of the optimized structures are calculated within the all-electron full-potential fully relativistic approach, with linear muffin-tin orbitals as basis functions, as implemented in the RSPt electronic structure code \cite{Wills1987,Wills2000,Wills2010}.

We have calculated magnetic interactions using the relativistic generalization \cite{Udvardi2003,Ebert2009,Kvashnin2020} of the Lichtenstein-Katsnelson-Antropov-Gubanov (LKAG) formula \cite{LKAG1987}, which we have recently applied to different systems \cite{Borisov2021}. This involves calculations of all components of the general interaction tensor $\hat{J}$ in the classical Heisenberg model:
\begin{equation}
    H = -\sum\limits_{i\neq j} J_{ij}^{\alpha\beta} e_i^\alpha e_j^\beta, \hspace{10pt} \alpha,\beta=x,y,z,
    \label{e:general_Heisenberg_model}
\end{equation}
where the unit vectors $\vec{e}_i$ indicate the direction of local spins and the exchange tensor $J_{ij}^{\alpha\beta}$ contains contributions from the Heisenberg exchange in the diagonal components as well as the DM interaction $\vec{D}$ and the symmetric anisotropic exchange $\hat{\Gamma}$ in the off-diagonal components:
\begin{equation}
    \hat{J}_{ij} = 
    \left(
    \begin{array}{ccc}
        J_{ij} & \Gamma^{xy}_{ij}+D^z_{ij} & \Gamma^{xz}_{ij}-D^y_{ij} \\[5pt]
        \Gamma^{xy}_{ij}-D^z_{ij} & J_{ij} & \Gamma^{yz}_{ij}+D^x_{ij} \\[5pt]
        \Gamma^{xz}_{ij}+D^y_{ij} & \Gamma^{yz}_{ij}-D^x_{ij} & J_{ij}
    \end{array}
    \right) \label{e:J_matrix}
\end{equation}
While the DM interaction is a subject of intense research, the symmetric anisotropic exchange $\hat{\Gamma}$ is seldom discussed in the literature, even though it can be important for the magnetic properties, as demonstrated in the present work. The calculations presented here in Section~3 are consistent with previous work \cite{Borisov2021} and indicate non-zero values of $\hat{\Gamma}$ for the B20 compounds, suggesting that this type of exchange can play an important role for the magnetic properties.

With the first-principles values of the magnetic interactions, the magnetic ground state is determined as a function of external magnetic field using micromagnetic simulations as implemented in the multiscale module $\mu$-\textsc{ASD} \cite{pereiro} of the \textsc{UppASD\-} code \cite{uppasd,Eriksson2017}, where the chosen size of the simulated region is $\unit[(500\times 500\times 100)]{nm}$. The micromagnetic energy density functional, derived starting from (\ref{e:general_Heisenberg_model}), reads:
\begin{equation}
    E[\vec{m}] = A\,(\vec{\nabla} \vec{m})^2 + \Gamma^{\alpha\beta}\vec{\nabla}m_\alpha\vec{\nabla}m_\beta + D\, \vec{m}\cdot(\vec{\nabla}\times\vec{m})
    \label{e:micromagnetic_energy}
\end{equation}
The effective micromagnetic parameters in this model, spin stiffness $A$, spiralization $D$ and symmetric exchange $\Gamma^{\alpha\beta}\:(\alpha,\beta=x,y,z)$, are determined from the $\mu \rightarrow 0$ limit of direct sums of the atomistic exchange parameters defined in Eqs.~(\ref{e:general_Heisenberg_model}) and (\ref{e:J_matrix}), using the following expressions:
\begin{equation}
  A = \frac{1}{2}\sum_{i\neq j} J_{ij}R^2_{ij}e^{-\mu R_{ij}}, \hspace{5pt} D = \sum_{i\neq j} (\vec{D}_{ij}\cdot \vec{R}_{ij}) e^{-\mu R_{ij}}
  \label{e:micromagnetic_parameters}
\end{equation}
\vspace{-10pt}
\begin{equation}
    \Gamma^{\alpha\beta} = \frac{1}{2}\sum_{i\neq j}\Gamma^{\alpha\beta}_{ij} R^2_{ij} e^{-\mu R_{ij}}
    \label{e:symmetric_exchange}
\end{equation}

The contribution of the symmetric anisotropic exchange $\Gamma^{\alpha\beta}$ to (\ref{e:micromagnetic_energy}) and expression (\ref{e:symmetric_exchange}) are derived by us for this work.

For 25\%-doped B20 compounds, all components $\Gamma^{\alpha\beta}$ are the same due to the lattice symmetry and the supercells considered in this work. The exponential factor in (\ref{e:micromagnetic_parameters}) and (\ref{e:symmetric_exchange}) is needed for the convergence of the sums with respect to the real-space cutoff radius for $R_{ij}$ (technical details in SI), and was also discussed for evaluations of spin-stiffness constant \cite{Pajda2001}.

In order to estimate the possibility of synthesis of the studied B20 compounds, we performed a structural stability analysis using the convex-hull method (workflow in Figure~\ref{f:convex_hull_workflow}). First, all the competing phases of the target B20 systems were determined by means of data mining with a data-driven high-throughput framework Python Materials Genomics (\textsc{Pymatgen}) \cite{ONG2013314} and only records with Inorganic Crystal Structure Database (ICSD) numbers in the Materials Project (MP) Database are selected. Then, the formation energy $\Delta U$ for each competing phase was calculated within DFT using open-source material information infrastructure AiiDA \cite{aiida} and Quantum Espresso (QE) software \cite{QE-2009,QE-2017} (further details in SI). The formation energy $\Delta U$ is defined as the difference between the total energy of the target compound and the total energy of the stoichiometric combination of its constituent pure elements in their standard ground states, e.g., ferromagnetic \textit{bcc} Fe and \textit{hcp} Co and non-magnetic Si (diamond structure). Based on the obtained $\Delta U$ values (Figure~\ref{f:formation_energy} in SI), the convex hull diagram (Figure~\ref{f:convex_hull_diagrams}) is constructed from all competing phases using the phase diagram method from \textsc{Pymatgen}. We also take into account the mixing entropy contribution $\Delta S$ to the free energy within the random-alloy approximation: $\Delta S = - k_\mathrm{B} [x \ln (x) + (1-x) \ln (1-x)]$, where $x$ is the dopant concentration. Since material synthesis usually takes place at elevated temperatures, we set $T=\unit[1000]{K}$ in our calculations, to get an estimate of the characteristic magnitude of $\Delta S$.

\section{Experimental Section}

\subsection{Synthesis}
Samples of Co$_{1-x}$Ru$_x$Si were synthesised by arc melting of stoichiometric quantities of Co (Goodfellow, purity 99.9\%), Ru (Cerac, purity 99.95\%) and Si (Goodfellow, purity 99.5\%) with a 5 at\% excess of Si. Oxidation was minimised by flushing  with Ar gas 5 times and melting a Ti getter before the sample synthesis. The samples were flipped and remelted 3 times to ensure good homogeneity. The sample was sealed under vacuum in a tantalum tube before annealing at $\unit[1773]{K}$ for 1 hour, slow cooling at $\unit[0.1]{K/min}$ to $\unit[1473]{K}$ and subsequently to room temperature. Single crystals were grown from the as cast samples which were crushed and heated in an induction furnace at $\unit[1673]{K}$ using the Bridgman method.

\subsection{Composition and structure analysis}
Powder X-ray diffraction data were measured on a Bruker D8 X-ray diffractometer equipped with a lynx-eye position sensitive detector using CuK$\alpha$ radiation ($\lambda = \unit[1.5418]{\AA}$). The sample was placed on a zero background single crystal silicon sample holder during data collection and the diffraction pattern was collected between 10 - 100$^{\circ}$ with a step size of 0.01$^{\circ}$. The data were analysed using Rietveld refinement \cite{Rietveld} within the topas6 software suite \cite{TOPAS}. Single crystal X-ray diffraction data were recorded on a Bruker D8 VENTURE diffractometer at $\unit[293]{K}$ using MoK$\alpha$\textsubscript{1} radiation ($\lambda = \unit[0.7107]{\AA}$). The data were processed using the APEX III software \cite{APEXIII} and subsequently solved and refined using the SHELX package within the WINGX program \cite{SHELXL,WinGX}. The composition of the compounds was investigated by scanning electron microscopy (SEM) using a Zeiss Leo 1550 field emission SEM equipped with an AZtec energy dispersive X-ray detector (EDS). The samples were prepared by grinding with SiC paper and subsequent polishing with SiO$\textsubscript{2}$ and H$\textsubscript{2}$O. Data were collected on at least 10 points using an accelerating voltage of 20 kV.

\medskip
\textbf{Acknowledgements} \par

This work was financially supported by the Knut and Alice Wallenberg Foundation through Grant No.\,2018.0060. O.E.~also acknowledges support by the Swedish Research Council (VR), the Foundation for Strategic Research (SSF), the Swedish Energy Agency (Energimyndigheten), the European Research Council (854843-FASTCORR), eSSENCE and STandUP. R.C. acknowledges support from the Swedish foundation for strategic Research (SSF) grant no.~EM-16-0039. J.C. acknowledges support from the VR grant no.~2019-00645. A.D. acknowledges support from the VR grants no.~2015-04608, 2016-05980 and 2019-05304. Q.X. acknowledges China Scholarship Council (201906920083). D.T. acknowledges support from the VR grant no.~2019-03666. The computations/data handling were enabled by resources provided by the Swedish National Infrastructure for Computing (SNIC) at the National Supercomputing Centre (NSC, Tetralith cluster), partially funded by the Swedish Research Council through grant agreement no.\,2018-05973. Structural sketches in Figure~\ref{f:D1_vectors} have been produced by the \textsc{VESTA3} software \cite{vesta}.

\medskip

\bibliographystyle{prb-titles.bst}
\bibliography{main}

\begin{thebibliography}{10}
\providecommand{\bibAnnoteFile}[1]{%
  \IfFileExists{#1}{\begin{quotation}\noindent\textsc{Key:} #1\\
  \textsc{Annotation:}\ \input{#1}\end{quotation}}{}}
\providecommand{\bibAnnote}[2]{%
  \begin{quotation}\noindent\textsc{Key:} #1\\
  \textsc{Annotation:}\ #2\end{quotation}}
\providecommand{\bibinfo}[2]{#2}

\bibitem{Aeppli1992}
\bibinfo{author}{G.~Aeppli} and \bibinfo{author}{Z.~Fisk},
  \bibinfo{journal}{Comments Condens. Matter Phys.}
  \textbf{\bibinfo{volume}{16}}, \bibinfo{pages}{155} (\bibinfo{year}{1992}).
\bibAnnoteFile{Aeppli1992}

\bibitem{Wernick1972}
\bibinfo{author}{J.~H. Wernick}, \bibinfo{author}{G.~K. Wertheim}, and
  \bibinfo{author}{R.~C. Sherwood}, \bibinfo{journal}{Mater. Res. Bull.}
  \textbf{\bibinfo{volume}{7}}, \bibinfo{pages}{1431} (\bibinfo{year}{1972}).
\bibAnnoteFile{Wernick1972}

\bibitem{Manyala2000}
\bibinfo{author}{N.~Manyala}, \bibinfo{author}{Y.~Sidis},
  \bibinfo{author}{J.~F. DiTusa}, \bibinfo{author}{G.~Aeppli},
  \bibinfo{author}{D.~Young}, and \bibinfo{author}{Z.~Fisk},
  \bibinfo{journal}{Nature} \textbf{\bibinfo{volume}{404}},
  \bibinfo{pages}{581} (\bibinfo{year}{2000}).
\bibAnnoteFile{Manyala2000}

\bibitem{Onose2005}
\bibinfo{author}{Y.~Onose}, \bibinfo{author}{N.~Takeshita},
  \bibinfo{author}{C.~Terakura}, \bibinfo{author}{H.~Takagi}, and
  \bibinfo{author}{Y.~Tokura}, \bibinfo{journal}{Phys. Rev. B}
  \textbf{\bibinfo{volume}{72}}, \bibinfo{pages}{224431}
  (\bibinfo{year}{2005}).
\bibAnnoteFile{Onose2005}

\bibitem{Muenzer2010}
\bibinfo{author}{W.~M\"unzer}, \bibinfo{author}{A.~Neubauer},
  \bibinfo{author}{T.~Adams}, \bibinfo{author}{S.~M\"uhlbauer},
  \bibinfo{author}{C.~Franz}, \bibinfo{author}{F.~Jonietz},
  \bibinfo{author}{R.~Georgii}, \bibinfo{author}{P.~B\"oni},
  \bibinfo{author}{B.~Pedersen}, \bibinfo{author}{M.~Schmidt},
  \bibinfo{author}{A.~Rosch}, and \bibinfo{author}{C.~Pfleiderer},
  \bibinfo{journal}{Phys. Rev. B} \textbf{\bibinfo{volume}{81}},
  \bibinfo{pages}{041203} (\bibinfo{year}{2010}).
\bibAnnoteFile{Muenzer2010}

\bibitem{Yu2010}
\bibinfo{author}{X.~Z. Yu}, \bibinfo{author}{Y.~Onose},
  \bibinfo{author}{N.~Kanazawa}, \bibinfo{author}{J.~H. Park},
  \bibinfo{author}{J.~H. Han}, \bibinfo{author}{Y.~Matsui},
  \bibinfo{author}{N.~Nagaosa}, and \bibinfo{author}{Y.~Tokura},
  \bibinfo{journal}{Nature} \textbf{\bibinfo{volume}{465}},
  \bibinfo{pages}{901} (\bibinfo{year}{2010}).
\bibAnnoteFile{Yu2010}

\bibitem{Muehlbauer2009}
\bibinfo{author}{S.~Mühlbauer}, \bibinfo{author}{B.~Binz},
  \bibinfo{author}{F.~Jonietz}, \bibinfo{author}{C.~Pfleiderer},
  \bibinfo{author}{A.~Rosch}, \bibinfo{author}{A.~Neubauer},
  \bibinfo{author}{R.~Georgii}, and \bibinfo{author}{P.~Böni},
  \bibinfo{journal}{Science} \textbf{\bibinfo{volume}{323}},
  \bibinfo{pages}{915} (\bibinfo{year}{2009}).
\bibAnnoteFile{Muehlbauer2009}

\bibitem{Dzyaloshinsky1958}
\bibinfo{author}{I.~Dzyaloshinsky}, \bibinfo{journal}{Journal of Physics and
  Chemistry of Solids} \textbf{\bibinfo{volume}{4}}, \bibinfo{pages}{241 }
  (\bibinfo{year}{1958}).
\bibAnnoteFile{Dzyaloshinsky1958}

\bibitem{Moriya1960}
\bibinfo{author}{T.~Moriya}, \bibinfo{journal}{Phys. Rev.}
  \textbf{\bibinfo{volume}{120}}, \bibinfo{pages}{91} (\bibinfo{year}{1960}).
\bibAnnoteFile{Moriya1960}

\bibitem{Heinze2011}
\bibinfo{author}{S.~Heinze}, \bibinfo{author}{K.~v. Bergmann},
  \bibinfo{author}{M.~Menzel}, \bibinfo{author}{J.~Brede},
  \bibinfo{author}{A.~Kubetzka}, \bibinfo{author}{R.~Wiesendanger},
  \bibinfo{author}{G.~Bihlmayer}, and \bibinfo{author}{S.~Bl\"ugel},
  \bibinfo{journal}{Nature Physics} \textbf{\bibinfo{volume}{7}},
  \bibinfo{pages}{713} (\bibinfo{year}{2011}).
\bibAnnoteFile{Heinze2011}

\bibitem{Wang2019}
\bibinfo{author}{L.~Wang}, \bibinfo{author}{C.~Liu},
  \bibinfo{author}{N.~Mehmood}, \bibinfo{author}{G.~Han},
  \bibinfo{author}{Y.~Wang}, \bibinfo{author}{X.~Xu},
  \bibinfo{author}{C.~Feng}, \bibinfo{author}{Z.~Hou},
  \bibinfo{author}{Y.~Peng}, \bibinfo{author}{X.~Gao}, and
  \bibinfo{author}{G.~Yu}, \bibinfo{journal}{ACS Appl. Mater. Interfaces}
  \textbf{\bibinfo{volume}{11}}, \bibinfo{pages}{12098} (\bibinfo{year}{2019}).
\bibAnnoteFile{Wang2019}

\bibitem{Soumyanarayanan2017}
\bibinfo{author}{A.~Soumyanarayanan}, \bibinfo{author}{M.~Raju},
  \bibinfo{author}{A.~L.~G. Oyarce}, \bibinfo{author}{A.~K.~C. Tan},
  \bibinfo{author}{M.-Y. Im}, \bibinfo{author}{A.~P. Petrović},
  \bibinfo{author}{P.~Ho}, \bibinfo{author}{K.~H. Khoo},
  \bibinfo{author}{M.~Tran}, \bibinfo{author}{C.~K. Gan},
  \bibinfo{author}{F.~Ernult}, and \bibinfo{author}{C.~Panagopoulos},
  \bibinfo{journal}{Nature Materials} \textbf{\bibinfo{volume}{16}},
  \bibinfo{pages}{898} (\bibinfo{year}{2017}).
\bibAnnoteFile{Soumyanarayanan2017}

\bibitem{Sidorov2018}
\bibinfo{author}{V.~A. Sidorov}, \bibinfo{author}{A.~E. Petrova},
  \bibinfo{author}{N.~M. Chtchelkatchev}, \bibinfo{author}{M.~V. Magnitskaya},
  \bibinfo{author}{L.~N. Fomicheva}, \bibinfo{author}{D.~A. Salamatin},
  \bibinfo{author}{A.~V. Nikolaev}, \bibinfo{author}{I.~P. Zibrov},
  \bibinfo{author}{F.~Wilhelm}, \bibinfo{author}{A.~Rogalev}, and
  \bibinfo{author}{A.~V. Tsvyashchenko}, \bibinfo{journal}{Phys. Rev. B}
  \textbf{\bibinfo{volume}{98}}, \bibinfo{pages}{125121}
  (\bibinfo{year}{2018}).
\bibAnnoteFile{Sidorov2018}

\bibitem{Dhital2017}
\bibinfo{author}{C.~Dhital}, \bibinfo{author}{L.~DeBeer-Schmitt},
  \bibinfo{author}{Q.~Zhang}, \bibinfo{author}{W.~Xie}, \bibinfo{author}{D.~P.
  Young}, and \bibinfo{author}{J.~F. DiTusa}, \bibinfo{journal}{Phys. Rev. B}
  \textbf{\bibinfo{volume}{96}}, \bibinfo{pages}{214425}
  (\bibinfo{year}{2017}).
\bibAnnoteFile{Dhital2017}

\bibitem{Sales1994}
\bibinfo{author}{B.~C. Sales}, \bibinfo{author}{E.~C. Jones},
  \bibinfo{author}{B.~C. Chakoumakos}, \bibinfo{author}{J.~A. Fernandez-Baca},
  \bibinfo{author}{H.~E. Harmon}, \bibinfo{author}{J.~W. Sharp}, and
  \bibinfo{author}{E.~H. Volckmann}, \bibinfo{journal}{Phys. Rev. B}
  \textbf{\bibinfo{volume}{50}}, \bibinfo{pages}{8207} (\bibinfo{year}{1994}).
\bibAnnoteFile{Sales1994}

\bibitem{pereiro}
\bibinfo{author}{E.~Méndez}, \bibinfo{author}{M.~Poluektov},
  \bibinfo{author}{G.~Kreiss}, \bibinfo{author}{O.~Eriksson}, and
  \bibinfo{author}{M.~Pereiro}, \bibinfo{journal}{Phys. Rev. Research}
  \textbf{\bibinfo{volume}{2}}, \bibinfo{pages}{013092} (\bibinfo{year}{2020}).
\bibAnnoteFile{pereiro}

\bibitem{uppasd}
\bibinfo{author}{B.~Skubic}, \bibinfo{author}{J.~Hellsvik},
  \bibinfo{author}{L.~Nordstr{\"o}m}, and \bibinfo{author}{O.~Eriksson},
  \bibinfo{journal}{Journal of Physics: Condensed Matter}
  \textbf{\bibinfo{volume}{20}}, \bibinfo{pages}{315203}
  (\bibinfo{year}{2008}).
\bibAnnoteFile{uppasd}

\bibitem{Eriksson2017}
\bibinfo{author}{O.~Eriksson}, \bibinfo{author}{A.~Bergman},
  \bibinfo{author}{L.~Bergqvist}, and \bibinfo{author}{J.~Hellsvik},
  \emph{\bibinfo{title}{Atomistic Spin Dynamics: Foundations and
  Applications}}, \bibinfo{publisher}{Oxford University Press, Oxford, UK}
  (\bibinfo{year}{2017}).
\bibAnnoteFile{Eriksson2017}

\bibitem{Perring1999}
\bibinfo{author}{L.~Perring}, \bibinfo{author}{F.~Bussy},
  \bibinfo{author}{J.~Gachon}, and \bibinfo{author}{P.~Feschotte},
  \bibinfo{journal}{J. Alloys Compd.} \textbf{\bibinfo{volume}{284}},
  \bibinfo{pages}{198} (\bibinfo{year}{1999}).
\bibAnnoteFile{Perring1999}

\bibitem{Hohenberg1964}
\bibinfo{author}{P.~Hohenberg} and \bibinfo{author}{W.~Kohn},
  \bibinfo{journal}{Phys. Rev.} \textbf{\bibinfo{volume}{136}},
  \bibinfo{pages}{B864} (\bibinfo{year}{1964}).
\bibAnnoteFile{Hohenberg1964}

\bibitem{Kohn1965}
\bibinfo{author}{W.~Kohn} and \bibinfo{author}{L.~J. Sham},
  \bibinfo{journal}{Phys. Rev.} \textbf{\bibinfo{volume}{140}},
  \bibinfo{pages}{A1133} (\bibinfo{year}{1965}).
\bibAnnoteFile{Kohn1965}

\bibitem{PBE1996}
\bibinfo{author}{J.~P. Perdew}, \bibinfo{author}{K.~Burke}, and
  \bibinfo{author}{M.~Ernzerhof}, \bibinfo{journal}{Phys. Rev. Lett.}
  \textbf{\bibinfo{volume}{77}}, \bibinfo{pages}{3865} (\bibinfo{year}{1996}).
\bibAnnoteFile{PBE1996}

\bibitem{Borisov2021}
\bibinfo{author}{V.~Borisov}, \bibinfo{author}{Y.~O. Kvashnin},
  \bibinfo{author}{N.~Ntallis}, \bibinfo{author}{D.~Thonig},
  \bibinfo{author}{P.~Thunstr\"om}, \bibinfo{author}{M.~Pereiro},
  \bibinfo{author}{A.~Bergman}, \bibinfo{author}{E.~Sj\"oqvist},
  \bibinfo{author}{A.~Delin}, \bibinfo{author}{L.~Nordstr\"om}, and
  \bibinfo{author}{O.~Eriksson}, \bibinfo{journal}{Phys. Rev. B}
  \textbf{\bibinfo{volume}{103}}, \bibinfo{pages}{174422}
  (\bibinfo{year}{2021}).
\bibAnnoteFile{Borisov2021}

\bibitem{Kresse1996}
\bibinfo{author}{G.~Kresse} and \bibinfo{author}{J.~Furthm\"uller},
  \bibinfo{journal}{Phys. Rev. B} \textbf{\bibinfo{volume}{54}},
  \bibinfo{pages}{11169} (\bibinfo{year}{1996}).
\bibAnnoteFile{Kresse1996}

\bibitem{Wills1987}
\bibinfo{author}{J.~M. Wills} and \bibinfo{author}{B.~R. Cooper},
  \bibinfo{journal}{Phys. Rev. B} \textbf{\bibinfo{volume}{36}},
  \bibinfo{pages}{3809} (\bibinfo{year}{1987}).
\bibAnnoteFile{Wills1987}

\bibitem{Wills2000}
\bibinfo{author}{J.~Wills}, \bibinfo{author}{O.~Eriksson},
  \bibinfo{author}{M.~Alouani}, and \bibinfo{author}{D.~Price},
  \emph{\bibinfo{title}{``Full-Potential LMTO Total Energy and Force
  Calculations'' in {\it Electronic structure and physical properties of
  solids}}}, \bibinfo{publisher}{Springer-Verlag Berlin Heidelberg}
  (\bibinfo{year}{2000}).
\bibAnnoteFile{Wills2000}

\bibitem{Wills2010}
\bibinfo{author}{J.~Wills}, \bibinfo{author}{M.~Alouani},
  \bibinfo{author}{P.~Andersson}, \bibinfo{author}{A.~Delin},
  \bibinfo{author}{O.~Eriksson}, and \bibinfo{author}{O.~Grechnyev},
  \emph{\bibinfo{title}{Full-Potential Electronic Structure Method}}, volume
  \bibinfo{volume}{167}, \bibinfo{publisher}{Springer-Verlag Berlin Heidelberg}
  (\bibinfo{year}{2010}).
\bibAnnoteFile{Wills2010}

\bibitem{Udvardi2003}
\bibinfo{author}{L.~Udvardi}, \bibinfo{author}{L.~Szunyogh},
  \bibinfo{author}{K.~Palot\'as}, and \bibinfo{author}{P.~Weinberger},
  \bibinfo{journal}{Phys. Rev. B} \textbf{\bibinfo{volume}{68}},
  \bibinfo{pages}{104436} (\bibinfo{year}{2003}).
\bibAnnoteFile{Udvardi2003}

\bibitem{Ebert2009}
\bibinfo{author}{H.~Ebert} and \bibinfo{author}{S.~Mankovsky},
  \bibinfo{journal}{Phys. Rev. B} \textbf{\bibinfo{volume}{79}},
  \bibinfo{pages}{045209} (\bibinfo{year}{2009}).
\bibAnnoteFile{Ebert2009}

\bibitem{Kvashnin2020}
\bibinfo{author}{Y.~O. Kvashnin}, \bibinfo{author}{A.~Bergman},
  \bibinfo{author}{A.~I. Lichtenstein}, and \bibinfo{author}{M.~I. Katsnelson},
  \bibinfo{journal}{Phys. Rev. B} \textbf{\bibinfo{volume}{102}},
  \bibinfo{pages}{115162} (\bibinfo{year}{2020}).
\bibAnnoteFile{Kvashnin2020}

\bibitem{LKAG1987}
\bibinfo{author}{A.~Liechtenstein}, \bibinfo{author}{M.~Katsnelson},
  \bibinfo{author}{V.~Antropov}, and \bibinfo{author}{V.~Gubanov},
  \bibinfo{journal}{Journal of Magnetism and Magnetic Materials}
  \textbf{\bibinfo{volume}{67}}, \bibinfo{pages}{65 } (\bibinfo{year}{1987}).
\bibAnnoteFile{LKAG1987}

\bibitem{Pajda2001}
\bibinfo{author}{M.~Pajda}, \bibinfo{author}{J.~Kudrnovsk\'y},
  \bibinfo{author}{I.~Turek}, \bibinfo{author}{V.~Drchal}, and
  \bibinfo{author}{P.~Bruno}, \bibinfo{journal}{Phys. Rev. B}
  \textbf{\bibinfo{volume}{64}}, \bibinfo{pages}{174402}
  (\bibinfo{year}{2001}).
\bibAnnoteFile{Pajda2001}

\bibitem{ONG2013314}
\bibinfo{author}{S.~P. Ong}, \bibinfo{author}{W.~D. Richards},
  \bibinfo{author}{A.~Jain}, \bibinfo{author}{G.~Hautier},
  \bibinfo{author}{M.~Kocher}, \bibinfo{author}{S.~Cholia},
  \bibinfo{author}{D.~Gunter}, \bibinfo{author}{V.~L. Chevrier},
  \bibinfo{author}{K.~A. Persson}, and \bibinfo{author}{G.~Ceder},
  \bibinfo{journal}{Computational Materials Science}
  \textbf{\bibinfo{volume}{68}}, \bibinfo{pages}{314} (\bibinfo{year}{2013}).
\bibAnnoteFile{ONG2013314}

\bibitem{aiida}
\bibinfo{author}{S.~P. Huber}, \bibinfo{author}{S.~Zoupanos},
  \bibinfo{author}{M.~Uhrin}, \bibinfo{author}{L.~Talirz},
  \bibinfo{author}{L.~Kahle}, \bibinfo{author}{R.~Häuselmann},
  \bibinfo{author}{D.~Gresch}, \bibinfo{author}{T.~Müller},
  \bibinfo{author}{A.~V. Yakutovich}, \bibinfo{author}{C.~W. Andersen}
  \emph{et~al.}, \bibinfo{journal}{Sci Data} \textbf{\bibinfo{volume}{7}},
  \bibinfo{pages}{300} (\bibinfo{year}{2020}).
\bibAnnoteFile{aiida}

\bibitem{QE-2009}
\bibinfo{author}{P.~Giannozzi}, \bibinfo{author}{S.~Baroni},
  \bibinfo{author}{N.~Bonini}, \bibinfo{author}{M.~Calandra},
  \bibinfo{author}{R.~Car}, \bibinfo{author}{C.~Cavazzoni},
  \bibinfo{author}{D.~Ceresoli}, \bibinfo{author}{G.~L. Chiarotti},
  \bibinfo{author}{M.~Cococcioni}, \bibinfo{author}{I.~Dabo} \emph{et~al.},
  \bibinfo{journal}{Journal of Physics: Condensed Matter}
  \textbf{\bibinfo{volume}{21}}, \bibinfo{pages}{395502 (19pp)}
  (\bibinfo{year}{2009}).
\bibAnnoteFile{QE-2009}

\bibitem{QE-2017}
\bibinfo{author}{P.~Giannozzi}, \bibinfo{author}{O.~Andreussi},
  \bibinfo{author}{T.~Brumme}, \bibinfo{author}{O.~Bunau},
  \bibinfo{author}{M.~B. Nardelli}, \bibinfo{author}{M.~Calandra},
  \bibinfo{author}{R.~Car}, \bibinfo{author}{C.~Cavazzoni},
  \bibinfo{author}{D.~Ceresoli}, \bibinfo{author}{M.~Cococcioni} \emph{et~al.},
  \bibinfo{journal}{Journal of Physics: Condensed Matter}
  \textbf{\bibinfo{volume}{29}}, \bibinfo{pages}{465901}
  (\bibinfo{year}{2017}).
\bibAnnoteFile{QE-2017}

\bibitem{Rietveld}
\bibinfo{author}{H.~M. Rietveld}, \bibinfo{journal}{Journal of Applied
  Crystallography} \textbf{\bibinfo{volume}{2}}, \bibinfo{pages}{65}
  (\bibinfo{year}{1969}).
\bibAnnoteFile{Rietveld}

\bibitem{TOPAS}
\bibinfo{author}{A.~A. Coelho}, \bibinfo{journal}{J. of Appl. Crystallogr}
  \textbf{\bibinfo{volume}{51}}, \bibinfo{pages}{210} (\bibinfo{year}{2018}).
\bibAnnoteFile{TOPAS}

\bibitem{APEXIII}
\bibinfo{journal}{Bruker AXS Inc., ``Bruker APEXIII'', Madison, Wisconsin, USA}
   (\bibinfo{year}{2012}).
\bibAnnoteFile{APEXIII}

\bibitem{SHELXL}
\bibinfo{author}{G.~M. Sheldrick}, \bibinfo{journal}{Acta Crystalogr. Sect. C}
  \textbf{\bibinfo{volume}{71}}, \bibinfo{pages}{3} (\bibinfo{year}{2015}).
\bibAnnoteFile{SHELXL}

\bibitem{WinGX}
\bibinfo{author}{L.~J. Farrugia}, \bibinfo{journal}{Journal of Applied
  Crystallography} \textbf{\bibinfo{volume}{45}}, \bibinfo{pages}{849}
  (\bibinfo{year}{2012}).
\bibAnnoteFile{WinGX}

\bibitem{vesta}
\bibinfo{author}{K.~Momma} and \bibinfo{author}{F.~Izumi},
  \bibinfo{journal}{Journal of Applied Crystallography}
  \textbf{\bibinfo{volume}{44}}, \bibinfo{pages}{1272} (\bibinfo{year}{2011}).
\bibAnnoteFile{vesta}

\bibitem{pereiro1}
\bibinfo{author}{R.~Yadav}, \bibinfo{author}{M.~Pereiro},
  \bibinfo{author}{N.~A. Bogdanov}, \bibinfo{author}{S.~Nishimoto},
  \bibinfo{author}{A.~Bergman}, \bibinfo{author}{O.~Eriksson},
  \bibinfo{author}{J.~van~den Brink}, and \bibinfo{author}{L.~Hozoi},
  \bibinfo{journal}{Phys. Rev. Materials} \textbf{\bibinfo{volume}{2}},
  \bibinfo{pages}{074408} (\bibinfo{year}{2018}).
\bibAnnoteFile{pereiro1}

\bibitem{Lejaeghereaad3000}
\bibinfo{author}{K.~Lejaeghere}, \bibinfo{author}{G.~Bihlmayer},
  \bibinfo{author}{T.~Bj{\"o}rkman}, \bibinfo{author}{P.~Blaha},
  \bibinfo{author}{S.~Bl{\"u}gel}, \bibinfo{author}{V.~Blum},
  \bibinfo{author}{D.~Caliste}, \bibinfo{author}{I.~E. Castelli},
  \bibinfo{author}{S.~J. Clark}, \bibinfo{author}{A.~Dal~Corso} \emph{et~al.},
  \bibinfo{journal}{Science} \textbf{\bibinfo{volume}{351}}
  (\bibinfo{year}{2016}).
\bibAnnoteFile{Lejaeghereaad3000}

\bibitem{Grigoriev2013}
\bibinfo{author}{S.~V. Grigoriev}, \bibinfo{author}{N.~M. Potapova},
  \bibinfo{author}{S.-A. Siegfried}, \bibinfo{author}{V.~A. Dyadkin},
  \bibinfo{author}{E.~V. Moskvin}, \bibinfo{author}{V.~Dmitriev},
  \bibinfo{author}{D.~Menzel}, \bibinfo{author}{C.~D. Dewhurst},
  \bibinfo{author}{D.~Chernyshov}, \bibinfo{author}{R.~A. Sadykov},
  \bibinfo{author}{L.~N. Fomicheva}, and \bibinfo{author}{A.~V. Tsvyashchenko},
  \bibinfo{journal}{Phys. Rev. Lett.} \textbf{\bibinfo{volume}{110}},
  \bibinfo{pages}{207201} (\bibinfo{year}{2013}).
\bibAnnoteFile{Grigoriev2013}

\bibitem{Weitzer1997}
\bibinfo{author}{F.~Weitzer}, \bibinfo{author}{L.~Perring},
  \bibinfo{author}{J.~Gachon}, \bibinfo{author}{P.~Feschotte}, and
  \bibinfo{author}{J.~Schuster}, \bibinfo{journal}{Proceedings -
  Electrochemical Society} \textbf{\bibinfo{volume}{39}}, \bibinfo{pages}{241}
  (\bibinfo{year}{1997}).
\bibAnnoteFile{Weitzer1997}

\bibitem{Demchenko2008}
\bibinfo{author}{P.~Demchenko}, \bibinfo{author}{J.~K\'{o}nczyk},
  \bibinfo{author}{O.~Bodak}, \bibinfo{author}{R.~Matvijishyn},
  \bibinfo{author}{L.~Muratova}, and \bibinfo{author}{B.~Marciniak},
  \bibinfo{journal}{Chem. Met. Alloys} \textbf{\bibinfo{volume}{1}},
  \bibinfo{pages}{50} (\bibinfo{year}{2008}).
\bibAnnoteFile{Demchenko2008}

\end{thebibliography}

\clearpage
\appendix

\setlength{\tabcolsep}{5pt}
\renewcommand{\arraystretch}{2}

{\centerline {\Large {\bf Supporting Information}}}

\setcounter{section}{0}
\section{Further calculation details}

In the presence of dopants, the optimized supercells show rhombohedral distortions, with the angles between the lattice vectors deviating from 90\textdegree{} by less than $\sim$0.3\textdegree. This is related to the fact that the supercell approach chosen in this work basically implies a periodic arrangement of impurities, which breaks certain symmetries, in this case, the cubic symmetry. Despite this fact, these calculations still allow to estimate the effect of alloying on the structural parameters, magnetic interactions and measurable quantities such as magnetization, spin stiffness and skyrmion size, as discussed in the main text and SI. In terms of predicted structural trends, we notice that the calculated lattice parameters of Fe$_{1-x}$Co$_x$Si and Fe$_{1-x}$Ir$_x$Si agree well with the measured ($\unit[4.48]{\AA}$) \cite{Muenzer2010} and interpolated ($\unit[4.54-4.55]{\AA}$) \cite{Sales1994} values from the literature. Also for the Co$_{1-x}$Ru$_x$Si series we find a good agreement between the theoretical prediction and measurements of the lattice parameter (Figure~\ref{fig:CoRuSi Unit Cells.jpg}).

\section{Atomistic exchange interactions}

In order to calculate all components of the magnetic exchange tensor $J_{ij}^{\alpha\beta}$ in the effective spin model (\ref{e:general_Heisenberg_model}), three independent calculations are performed for each system assuming the ferromagnetic alignment of the transition metal moments along the $x$-, $y$- and $z$-directions. This allows to extract the three components of the DM vectors and symmetric anisotropic exchange $\Gamma^{\alpha\beta}$, both of which contribute to the exchange tensor defined by (\ref{e:J_matrix}). Due to the almost cubic symmetry and low magnetic anisotropy of the studied B20 compounds, the Heisenberg exchange parameters obtained from these independent calculations are nearly identical, which is in contrast to highly anisotropic systems, such as multilayers and surfaces, where one can expect a sizeable dependence of the Heisenberg exchange on the magnetic ordering axis. Low smearing of $\unit[1]{mRy}$ for electronic occupations and fine $(20\times20\times20)$ $k$-sampling of the Brillouin zone were used to ensure the convergence of the calculated magnetic interactions.

\section{Monte Carlo simulations and magnon spectra}

\begin{figure}
\centering
\includegraphics[width=0.99\columnwidth]{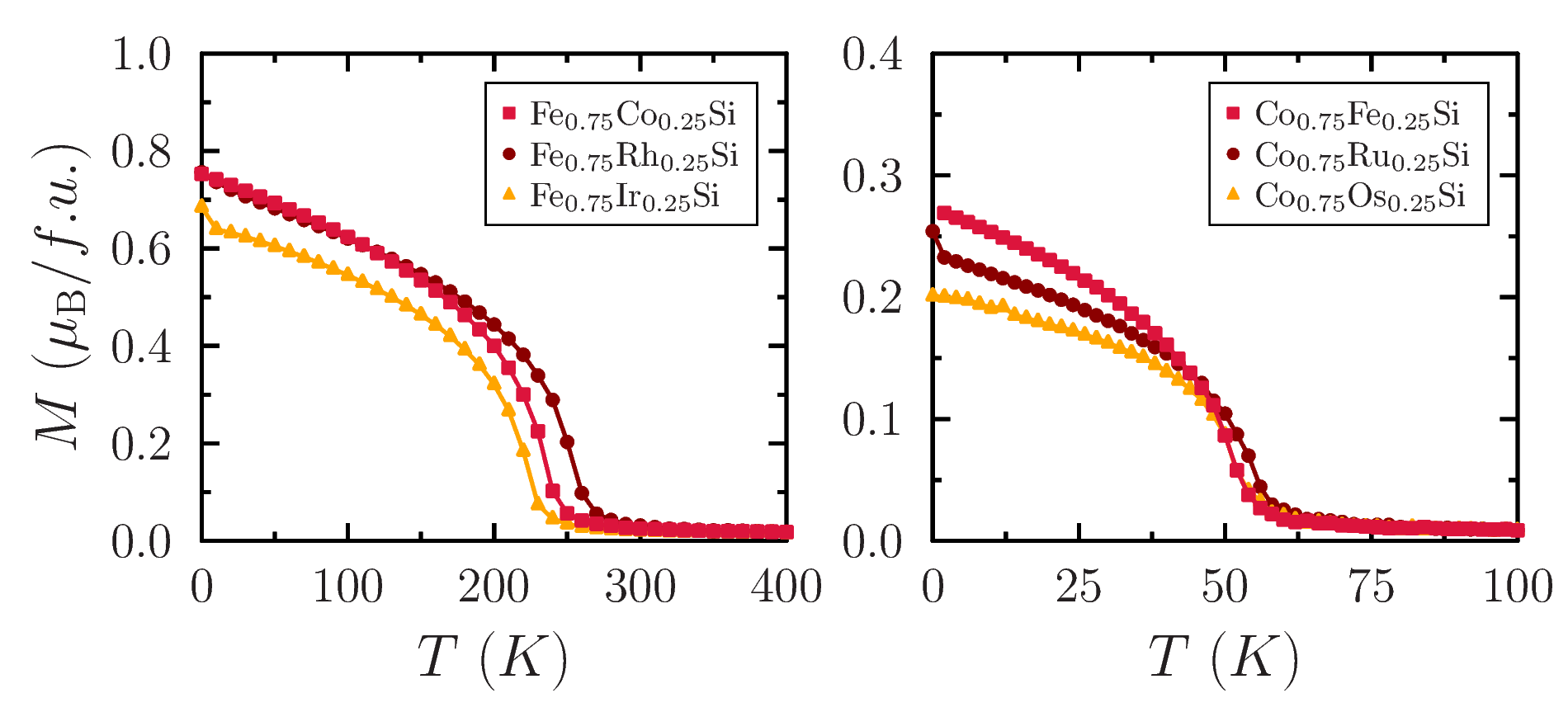}
\caption{Magnetic moment within the muffin tins per formula unit plotted versus temperature for the B20 compounds {\FS} (\textit{TM}=Co, Rh, Ir) and {\CS} ({\textit{TM}=Fe, Ru, Os}). The zero-temperature values are slightly different compared to Table~1, which is due to the contribution of the interstitial spin density.}
\label{fig:magnetization}
\end{figure}

In this section, we discuss the measurable magnetic properties of the doped B20 compounds {\FS} (\textit{TM}=Co, Rh, Ir) and {\CS} (\textit{TM}=Fe, Ru, Os) while the trends in the Fe$_{1-x}$Co$_x$Si and Co$_{1-x}$Ru$_x$Si series are discussed in section~\ref{SI_magnetic}. In Figure~\ref{fig:magnetization}, the temperature-dependence of the total magnetization $M(T)$ has similar shape for all systems, showing a quasi-linear decrease even at low temperatures, which is due to the classical nature of these Monte Carlo simulations. The magnetic moment per formula unit at $T=\unit[0]{K}$ (Table~\ref{tab:ADM}) is almost identical for the 3\textit{d} and 4\textit{d}-doped compounds, but it is reduced by 10\% for the 5\textit{d}-doped compounds. The Fe moment in the B20 compounds with Co, Rh and Ir dopants shows moderate variations and equals $\unit[0.88]{\mu_\mathrm{B}}$, $\unit[0.97]{\mu_\mathrm{B}}$ and $\unit[0.76]{\mu_\mathrm{B}}$, respectively. In the Co-based compounds, the Co moment shows larger variations and equals $\unit[0.21]{\mu_\mathrm{B}}$, $\unit[0.31]{\mu_\mathrm{B}}$ and $\unit[0.29]{\mu_\mathrm{B}}$ for the Fe-, Ru- and Os-doped systems, respectively. At the same time, the total magnetization is the largest for Co$_{0.75}$Fe$_{0.25}$Si due to the significant contribution from Fe sites, while the induced moments on Ru and Os in other compounds are much smaller due to the extended 4\textit{d} and 5\textit{d} orbitals. Lower Fe moment in 3\textit{d}-doped systems can be attributed to the fact that the magnetization is distributed among Fe and the dopant atoms, while in the 4\textit{d}- and 5\textit{d}-doped systems magnetic moments are mainly present on Fe.

\begin{figure}
\centering
\includegraphics[width=0.99\columnwidth]{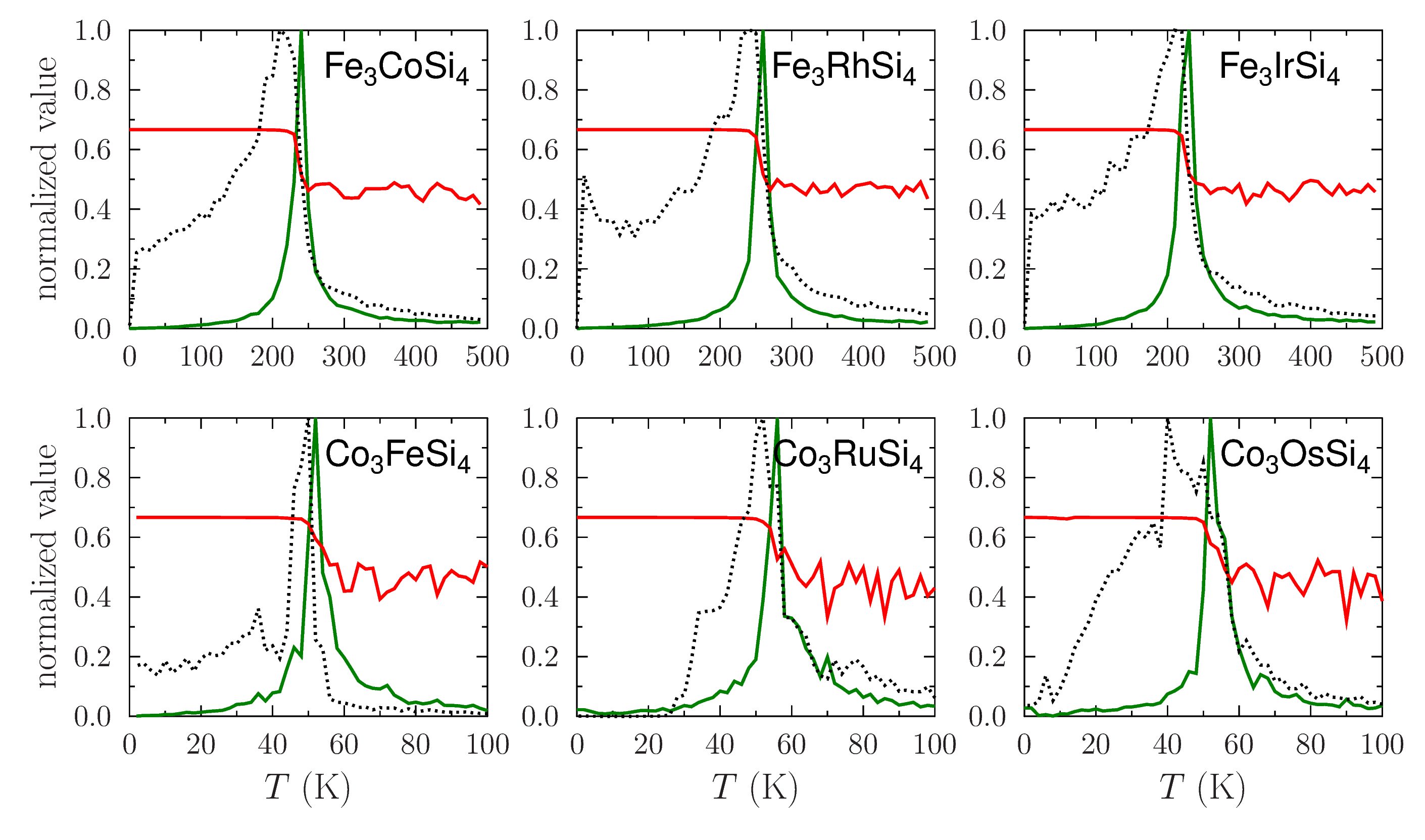}
\caption{Binder cumulants (red curve) and normalized susceptibility (green curve) and specific heat (black dotted curve) vs temperature from Monte Carlo simulations for the B20 compounds {\FS} (\textit{TM}=Co, Rh, Ir) and {\CS} ({\textit{TM}=Fe, Ru, Os}). The Curie temperature, determined from the peak of the susceptibility, and the zero-temperature magnetization are summarized in Table~\ref{tab:ADM}.}
\label{fig:cumulants}
\end{figure}

The Curie temperature $T_c$ is estimated from the temperature dependence of the magnetization and susceptibility calculated using the Monte Carlo technique applied to the generalized Heisenberg model (Equation~\ref{e:general_Heisenberg_model}) for a system of at least $(25\times 25\times 25)$ unit cells. The magnon spectra have been determined within a linear spin-wave theory (LSWT) framework in the adiabatic approximation using all the \textit{ab initio} magnetic interactions\cite{pereiro1}, including the symmetric anisotropic exchange $\Gamma^{\alpha\beta}$ in Equation~\ref{e:J_matrix}.

\begin{figure}
\begin{centering}
\includegraphics[width=0.99\columnwidth]{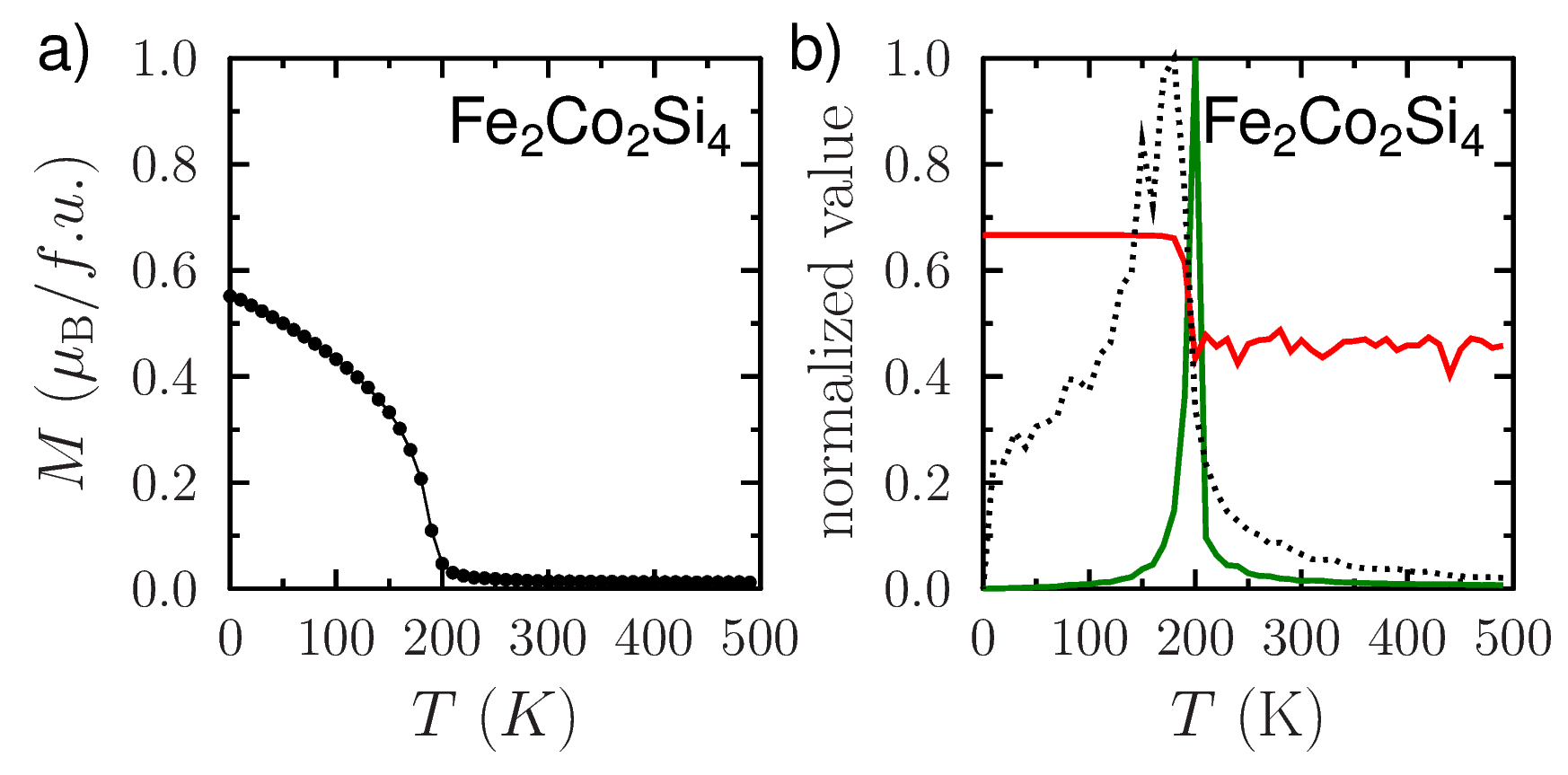}
\end{centering}
\vspace{-10pt}
\caption{a) Magnetic moment per unit cell plotted versus the temperature from Monte Carlo simulations for the B20 compound Fe$_{0.50}$Co$_{0.50}$Si. b) Binder cumulant (red curve), susceptibility $\chi$ (green curve) and specific heat C$_v$ (black dotted curve) as a function of the reduced temperature $T/T_c$ for the same compound.}
\label{fig:magnetization_supp}
\end{figure}

The magnetic ordering temperature $T_c$ of both Fe- and Co-based series is moderately changed by doping, which is evidenced by the thermomagnetic curves (Figure~\ref{fig:magnetization}). A more accurate estimation of $T_c$ (Table~\ref{tab:ADM}) is based on the following physical properties: Binder cumulants, magnetic susceptibility and specific heat which we obtain from the Monte Carlo simulations and plot in Figure~\ref{fig:cumulants} and \ref{fig:magnetization_supp}. The peaks in the susceptibility and specific heat agree with each other and are in a good agreement with the abrupt transition of the Binder cumulants, which all indicate the ordering temperature $T_c$. In general, the Fe-based compounds have higher $T_c$ than the Co-based compounds, which is due to larger magnetic moments (Table~\ref{tab:ADM}) and stronger exchange interactions (Figure~\ref{f:Jij_FeTMSi} and \ref{f:Jij_CoTMSi}).

An important comment should be made here. The overall magnitude of the calculated magnetic moments of the B20 compounds is rather overestimated, at least, compared to the data for the known Fe$_{1-x}$Co$_x$Si compounds \cite{Onose2005}. This leads to overestimation of the magnetic interactions in Figure~\ref{f:Jij_FeTMSi} and \ref{f:Jij_CoTMSi} and the ordering temperature $T_c$ in Table~1 in SI. Despite these deficiencies, we believe that our calculations still provide useful and reliable trends across both doping series, for example, for the $D/A$ ratio which determines the skyrmion size. The reason is that both the spiralization $D$ and spin stiffness $A$ originate in bilinear interactions described by Equation~(\ref{e:general_Heisenberg_model}) and (\ref{e:J_matrix}) and are expected to scale in the same way as functions of the magnetic moments.

\begin{figure}
\centering
\includegraphics[width=0.99\columnwidth]{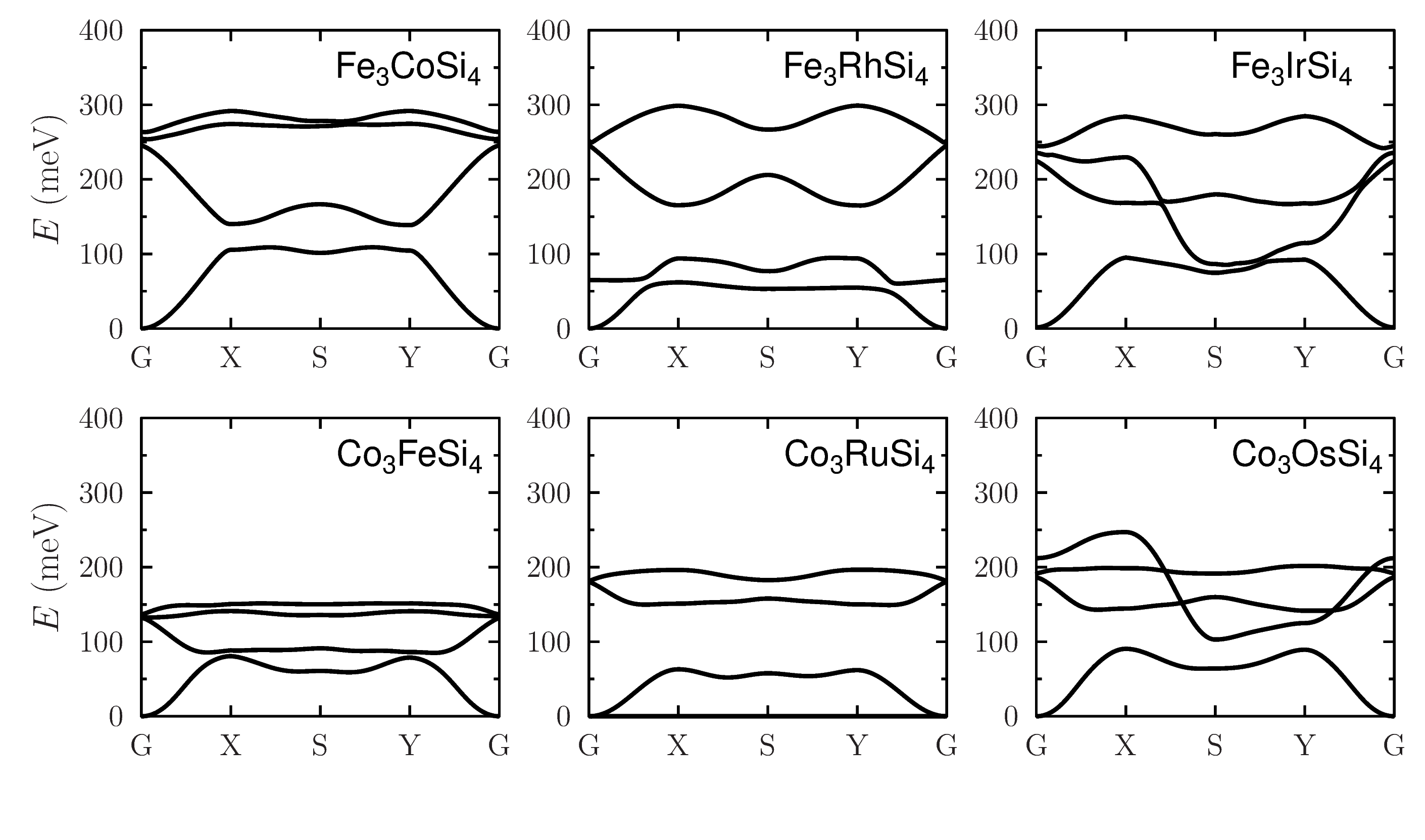}
\caption{Adiabatic magnon spectra for the doped B20 compounds calculated using linear spin-wave theory along the high symmetry points of the cubic Brillouin zone. In the first row, the spectra are shown for {\FS} (\textit{TM}=Co, Rh, Ir) while the second row shows {\CS} (\textit{TM}=Fe, Ru, Os).}
\label{fig:ams}
\end{figure}

\begin{figure}
\centering
\includegraphics[width=0.9\columnwidth]{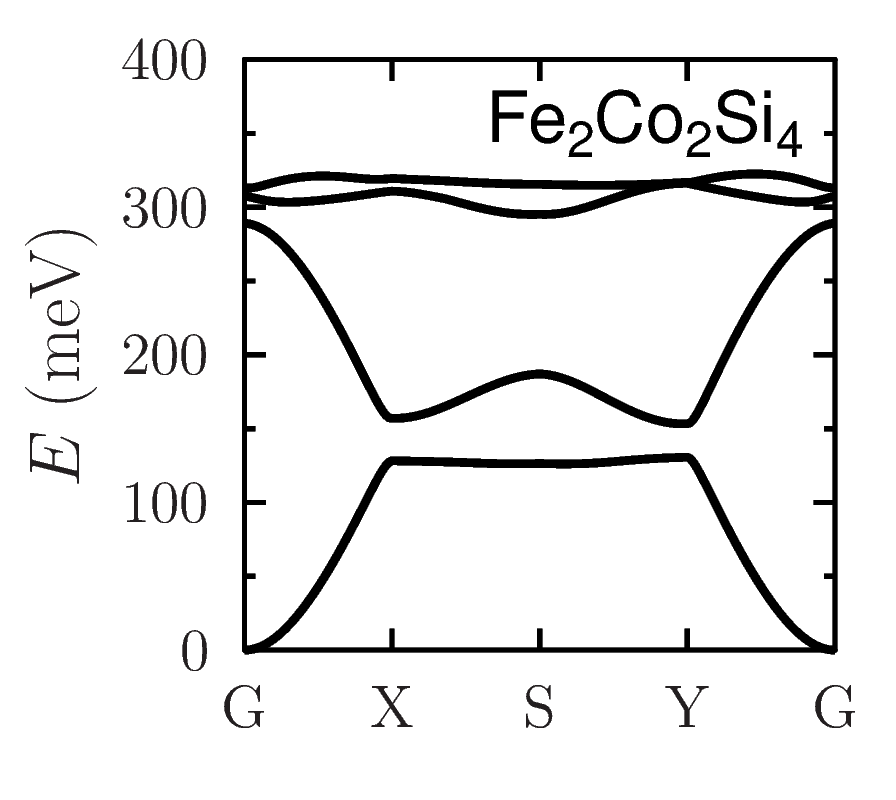}
\caption{Adiabatic magnon spectrum of Fe$_{0.50}$Co$_{0.50}$Si calculated using linear spin wave theory.}
\label{fig:ams_FeCoSi_supp}
\end{figure}

The calculated adiabatic magnon spectra of doped B20 compounds are shown in Figure~\ref{fig:ams} and \ref{fig:ams_FeCoSi_supp}. The spectra show four magnon branches, which is a consequence of the number of magnetic atoms per unit cell included in the spin model (\ref{e:general_Heisenberg_model}). The acoustic branch around the $\Gamma$ point shows a parabolic profile corroborating the ferromagnetic state of these compounds and its curvature along the $\Gamma-X$ direction is proportional to the exchange stiffness ($A$ values in Table~\ref{tab:ADM}). It is worthwhile to mention here that, from a topological point of view, the most interesting systems are Fe$_{0.75}$Co$_{0.25}$Si, Co$_{0.75}$Fe$_{0.25}$Si and Co$_{0.75}$Os$_{0.25}$Si because of the flat magnon band appearing in the acoustic branch along the directions $X-S-Y$. Since the inertial effects introduced by the mass of magnons are inversely proportional to the $\partial \rm{E}/\partial \textbf{q}$ where $\textbf{q}$ is the wave-vector, then the spin-wave velocity of the excited magnons at the energies given in Figure~\ref{fig:ams} ($\sim\unit[40]{meV}$ for Co$_{0.75}$Fe$_{0.25}$Si, $\sim\unit[90]{meV}$ for Fe$_{0.75}$Co$_{0.25}$Si and $\sim\unit[22]{meV}$ for Co$_{0.75}$Os$_{0.25}$Si) are significantly smaller, since the mass tends to infinity. This characteristic makes the aforementioned three compounds interesting for possible spintronics applications where magnons have to be localized in space.

\setlength{\tabcolsep}{5pt}
\renewcommand{\arraystretch}{1.5}

\begin{table*}
 \caption{Optimized lattice parameter $a$ together with the atomic magnetic moments of the main element ($m_1$) and dopant ($m_2$) are listed for the studied B20 compounds Fe$_{0.75}$\textit{TM}$_{0.25}$Si (\textit{TM}=Co, Rh, Ir) and Co$_{0.75}$\textit{TM}$_{0.25}$Si ({\textit{TM}=Fe, Ru, Os}). The ratio between the average nearest-neighbor DM ($D_1$) and Heisenberg interactions ($J_1$) is provided as well. In the right part of the table, calculated values of the exchange stiffness $A$, the spiralization $D$ and the symmetric anisotropic exchange $\Gamma^{\alpha\beta}$ (further information in the caption of Figure~\ref{f:micromagnetic_convergence}) defined by Equation~(\ref{e:micromagnetic_parameters}) and (\ref{e:symmetric_exchange}) as well as the $|D/A|$ ratio are shown along with the Curie temperature $T_c$ and the total spin magnetic moment $M$ (per unit cell with four magnetic sites), including the interstitial contribution.
 }
 \vspace{5pt}
 \centering
 \begin{tabular}{l|cccc|ccccccc}
    \hline
    Compound & $a$ & $m_1$ & $m_2$ & $\langle\nicefrac{D_1}{J_1}\rangle$  &      $A$        &         $D$   & $\Gamma^{\alpha\beta}$      & $|D/A|$ & $T_c$ & $M$ \\ [-5pt]
             & \AA & $\mu_\mathrm{B}$ & $\mu_\mathrm{B}$ & (\%) & $\mathrm{meV}\cdot\mathrm{\AA}^2$ & $\mathrm{meV}\cdot\mathrm{\AA}$  & $\mathrm{meV}\cdot\mathrm{\AA}^2$ & \AA$^{-1}$  &  $K$  & $\mu_\mathrm{B}/\mathrm{f.u.}$ \\
    \hline
    Fe$_{0.75}$Co$_{0.25}$Si  &  4.453 & 0.768 & 0.349 & 7.1 &  214.4  &  $-0.43$  &  +0.16 & 0.0020 & 240 & 0.702   \\
    Fe$_{0.75}$Rh$_{0.25}$Si  &  4.541 & 0.871 & 0.087 & 5.0-8.4 &  236.4  &  $-0.46$ & $-0.04$  & 0.0019 & 260 & 0.702   \\
    Fe$_{0.75}$Ir$_{0.25}$Si  &  4.553 & 0.755 & 0.060 & 10.6-11.2 & 224.6  &  $-0.62$ & $-1.29$ & 0.0028  & 230 & 0.640   \\
    \hline
    Co$_{0.75}$Fe$_{0.25}$Si  &  4.437 & 0.183 & 0.386 & 6.5 & 90.4 &   +0.38  & $-0.116$ & 0.0042  & 52 & 0.259    \\
    Co$_{0.75}$Ru$_{0.25}$Si  &  4.523 & 0.255 & 0.066 & 8.3 &  81.2  &   $-0.56$  & $-0.29$ & 0.0069 & 56 & 0.252    \\
    Co$_{0.75}$Os$_{0.25}$Si  &  4.538 & 0.210 & 0.046 & 10.3 & 71.2  &   $-1.09$  & $-2.02$ & 0.0153 & 52 & 0.233   \\
    \hline
 \end{tabular}
    \label{tab:ADM}
\end{table*}

The calculated spectra of Fe$_{0.75}$Ir$_{0.25}$Si in Figure~\ref{fig:ams} show a small acoustic magnon energy gap around $\unit[1.5]{meV}$, which is in contrast to other studied compounds and is due to the symmetric anisotropic exchange interaction~(\ref{e:symmetric_exchange}). This is confirmed by comparing the adiabatic magnon spectra calculated with or without this interaction. In both cases, the overall profiles of the spectra are basically the same but the magnon energy gap becomes zero when the symmetric anisotropic exchange is disregarded.

\newpage
\section{Micromagnetic parameters}

\begin{figure*}
\begin{centering}
\includegraphics[width=0.99\textwidth]{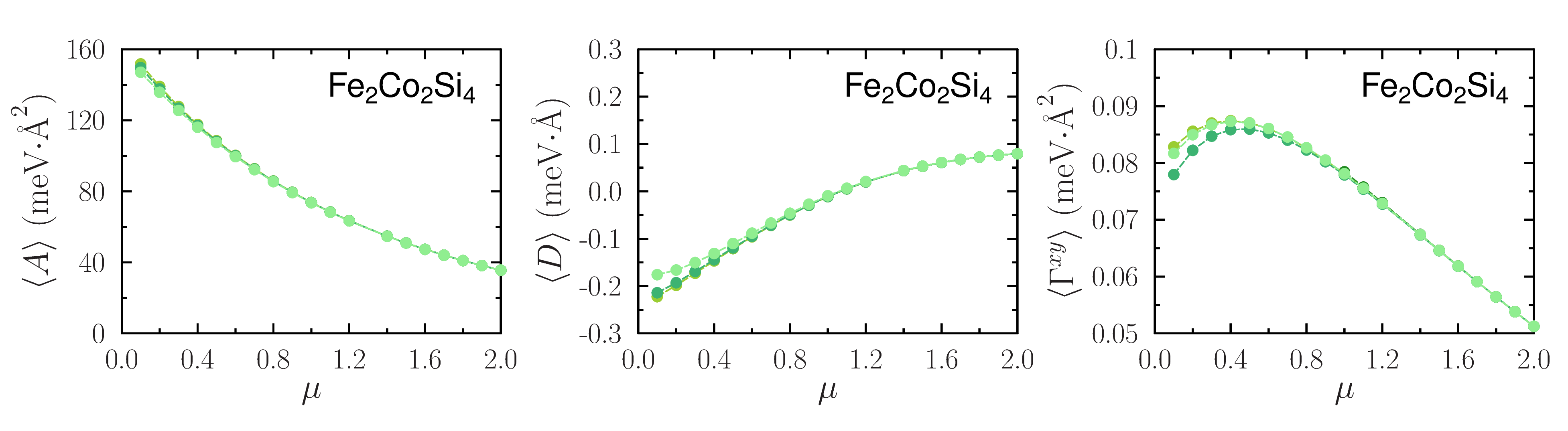}
\includegraphics[width=0.99\textwidth]{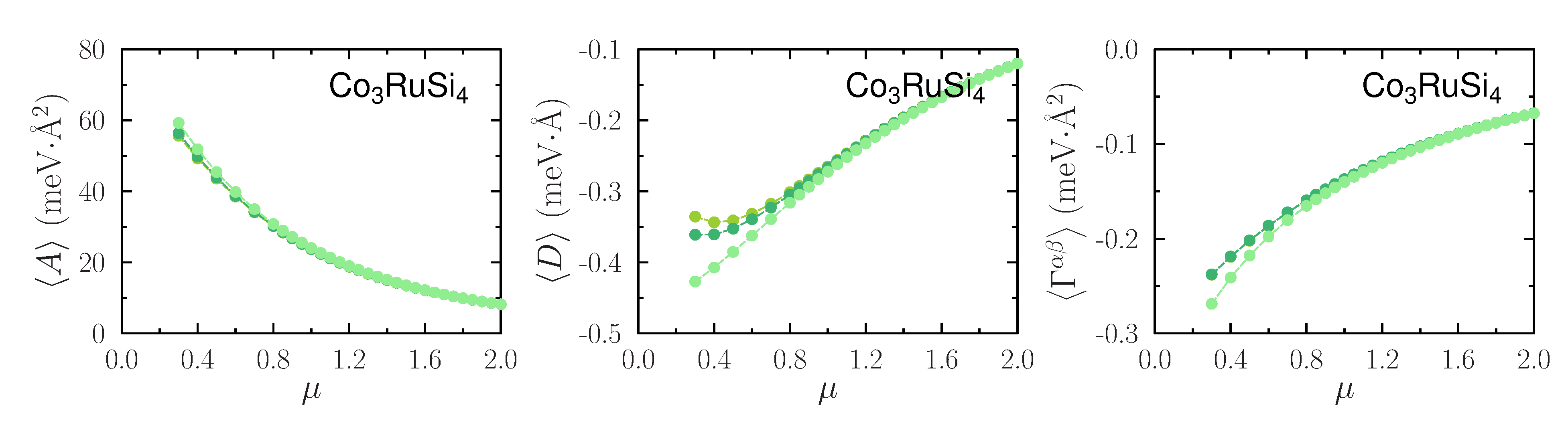}
\includegraphics[width=0.99\textwidth]{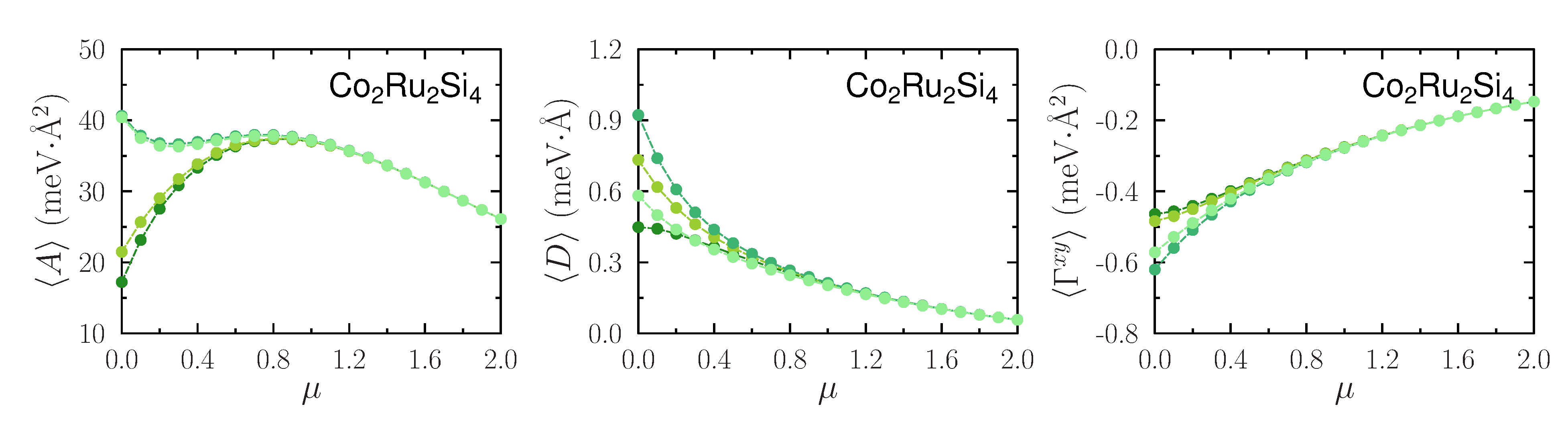}
\includegraphics[width=0.99\textwidth]{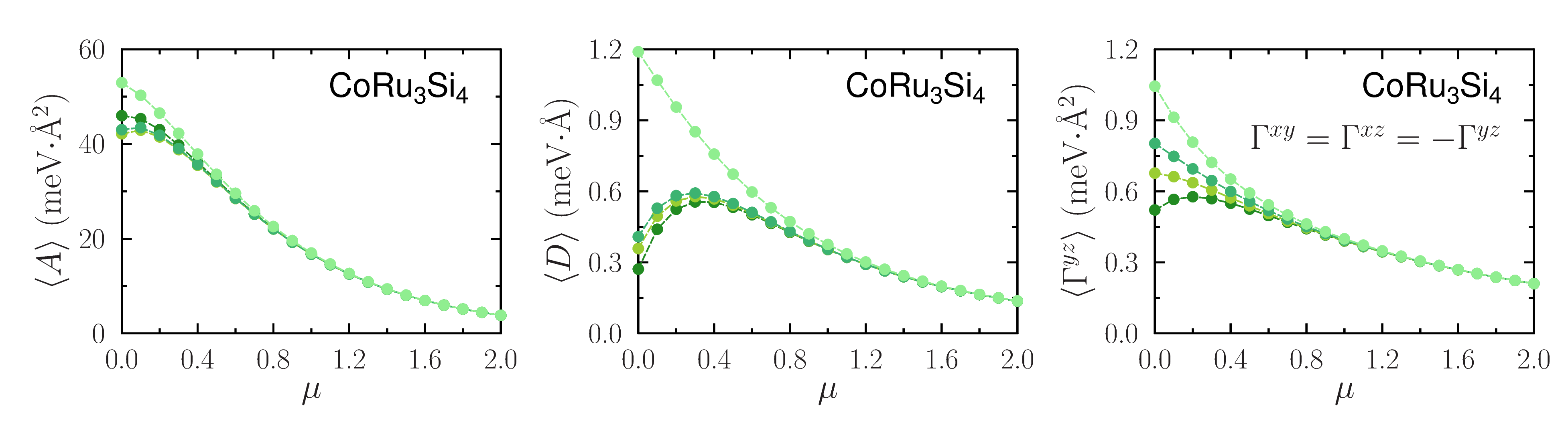}
\end{centering}
\caption{Dependence of the micromagnetic parameters of Fe$_{0.5}$Co$_{0.5}$Si, Co$_{0.75}$Ru$_{0.25}$Si and Co$_{0.25}$Ru$_{0.75}$Si on the $\mu$ parameter from the prefactor $e^{-\mu R_{ij}}$ which improves the convergence of sums in Equation~(\ref{e:micromagnetic_parameters}) and (\ref{e:symmetric_exchange}). Different curves in the same plot correspond to different real-space cutoffs for the distance $R_{ij}$, where darker color means larger $R_{ij}$ (up to around five lattice constants). For Co$_3$RuSi$_4$, all components of the symmetric exchange are the same ($\Gamma^{xy}=\Gamma^{xz}=\Gamma^{yz}$), while $\Gamma^{xz}=\Gamma^{yz}=0$ for Co$_2$Ru$_2$Si$_4$ and Fe$_2$Co$_2$Si$_4$, and $\Gamma^{xy}=\Gamma^{xz}=-\Gamma^{yz}$ for CoRu$_3$Si$_4$.}
\label{f:micromagnetic_convergence}
\end{figure*}
\begin{figure}
\centering
\includegraphics[width=0.99\columnwidth]{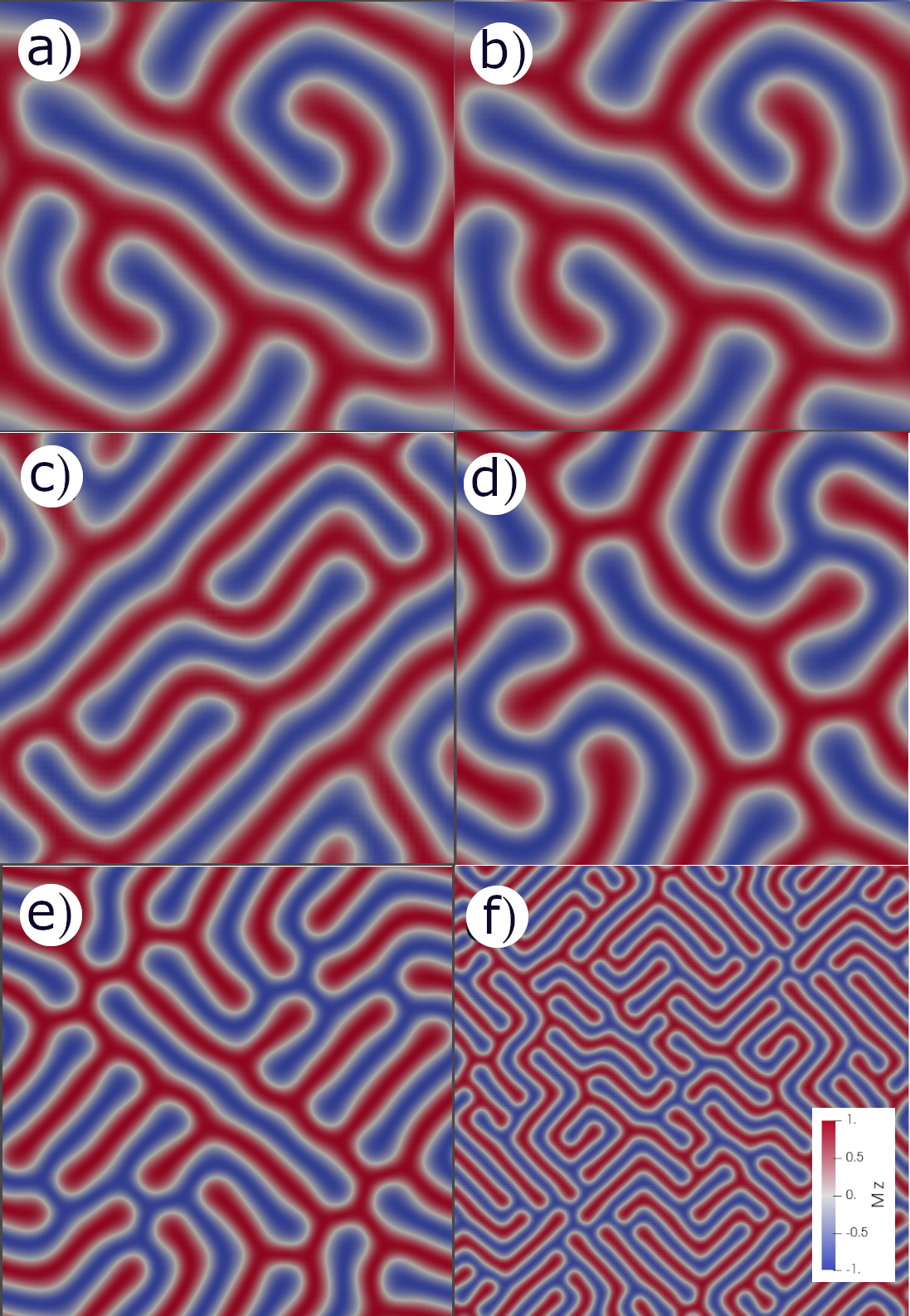}
\caption{Fragment of the magnetic ground state under zero magnetic field calculated with the multiscale module $\mu$-\textsc{ASD} for a) Fe$_{0.75}$Co$_{0.25}$Si b) Fe$_{0.75}$Rh$_{0.25}$Si, c)  Fe$_{0.75}$Ir$_{0.25}$Si, d) Co$_{0.75}$Fe$_{0.25}$Si e) Co$_{0.75}$Ru$_{0.25}$Si and f) Co$_{0.75}$Os$_{0.25}$Si. The simulation box size for the Fe-based compounds is $(3\times3\times1)\,\mu$m while it is $(1\times1\times0.3)\,\mu$m for the Co-based compounds. The color indicates the $z$-component of the magnetic moment, such that the red (blue) color represents the moments pointing fully along the $+z$-direction ($-z$-direction).}
\label{fig:magnetic_textures}
\end{figure}

For micromagnetic simulations of spin-spirals and skyrmions it is necessary to obtain effective parameters in Equation~(\ref{e:micromagnetic_energy}) according to expressions (\ref{e:micromagnetic_parameters}) and (\ref{e:symmetric_exchange}). In principle, we are interested here in the limit $\mu=0$ of these expressions which corresponds to the actual definition of these parameters. However, the numerical evaluation of these sums is difficult due to a slow convergence of the results with respect to the real-space cutoff radius for the atomistic interactions. The later decay with distance, of course, but since the interactions are multiplied with the number of equivalent neighbors and distance or distance squared, the sum converges rather slowly, especially for the DM interaction $D$ and symmetric anisotropic exchange $\Gamma^{\alpha\beta}$. When the regularization factor $e^{-\mu R_{ij}}$ is introduced, as suggested in a previous work \cite{Pajda2001}, the convergence is improved but the result does not always equal the $\mu=0$ limit. In order to estimate the later, we look at the $\mu$-dependence of the calculated $A$, $D$ and $\Gamma^{\alpha\beta}$ parameters for different cutoff radii (example in Figure~\ref{f:micromagnetic_convergence}) and find the region of $\mu$-values where different curves coincide (usually, for $\mu\geq 1.0-1.2$). These data are used then for fitting to the exponential function $f(\mu)=a\cdot e^{-b\mu} + c$ and calculating the $\mu=0$ limit as $f(0)=a+c$. The estimates of $A$, $D$ and $\Gamma^{\alpha\beta}$ obtained in this way are listed in Table~1 and are the basis of the micromagnetic simulations shown in Figure~\ref{fig:micromagnetic}, \ref{fig:magnetic_textures} and \ref{fig:Fe2Co2Si4_texture}.

\begin{table}
 \caption{Theoretical predictions for the spiral wave-length $L$ and skyrmion diameter $d$ for the doped B20 compounds {\FS} (\textit{TM}=Co, Rh, Ir), {\CS} ({\textit{TM}=Fe, Ru, Os}), Co$_{0.50}$Fe$_{0.50}$Si, Co$_{0.50}$Ru$_{0.50}$Si and Co$_{0.25}$Ru$_{0.75}$Si. The spiral wave-length is shown at zero magnetic field. The skyrmion diameter $d$ is provided at a field strength $B$ shown in the last column.}
\vspace{2pt}
  \centering
  \begin{tabular}{l|ccc}
    \hline
    Compound & $L$ [nm] & $d$ [nm] & $B$ [mT] \\
    \hline
    Fe$_{0.75}$Co$_{0.25}$Si  &  312&148&3\\
    Fe$_{0.75}$Rh$_{0.25}$Si  &  357&137&3\\
    Fe$_{0.75}$Ir$_{0.25}$Si  &  339&131&5\\
    \hline
    Co$_{0.75}$Fe$_{0.25}$Si  &  200&85&20\\
    Co$_{0.75}$Ru$_{0.25}$Si  &  99&130&20\\
    Co$_{0.75}$Os$_{0.25}$Si  &  84&50&200\\
    \hline
    Co$_{0.50}$Fe$_{0.50}$Si  &  191&80&3\\
    Co$_{0.50}$Ru$_{0.50}$Si  &  79&60&20\\
    Co$_{0.25}$Ru$_{0.75}$Si  &  128&70&40\\
    \hline
  \end{tabular}
  \label{tab:micro_param}
\end{table}

\newpage
\section{Stability analysis}

\begin{figure}
\centering
\includegraphics[width=0.99\columnwidth]{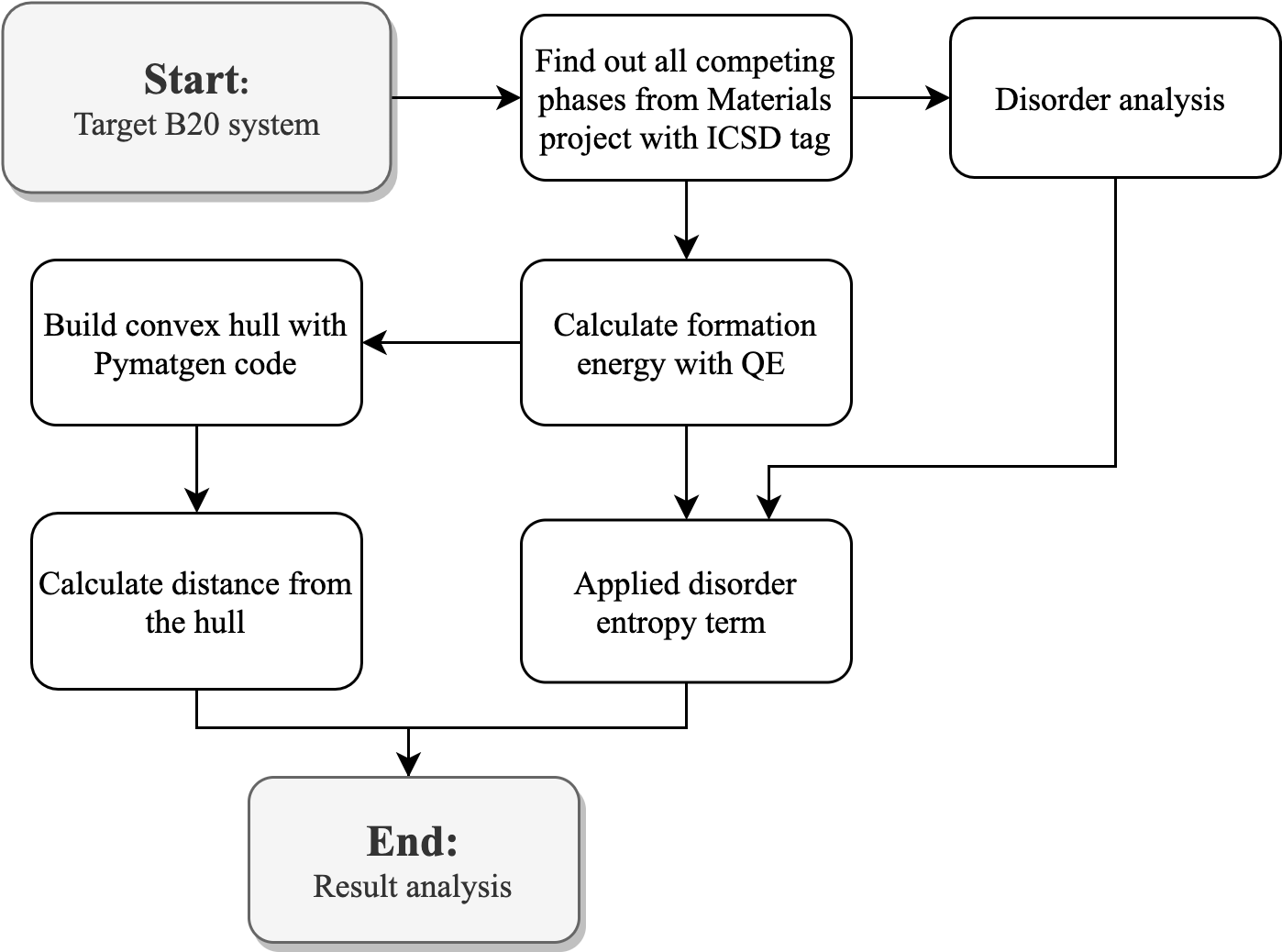}
\caption{Schematic representation of the workflow for the structural stability analysis in Sec.~III\,C.}
\label{f:convex_hull_workflow}
\end{figure}

\begin{figure}
\centering
\includegraphics[width=0.99\columnwidth]{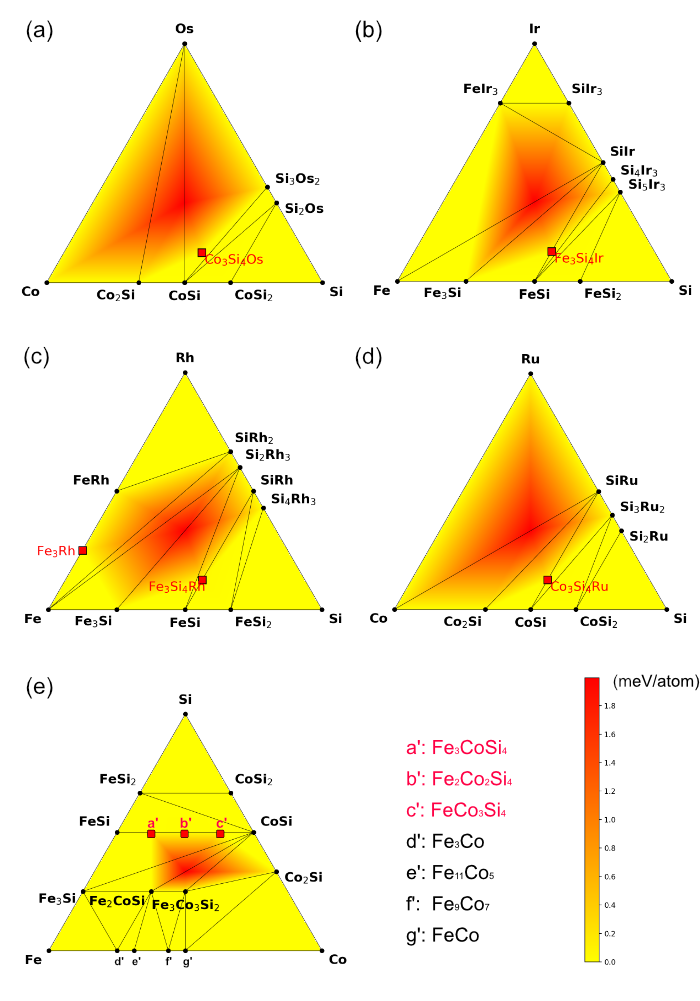}
\caption{Convex-hull diagrams for a) Co-Os-Si, b) Fe-Ir-Si, c) Fe-Rh-Si, d) Co-Ru-Si, e) Fe-Co-Si. The color code shows the distance from the convex hull.}
\label{f:convex_hull_diagrams}
\end{figure}

\begin{figure}
\centering
\includegraphics[width=0.99\columnwidth]{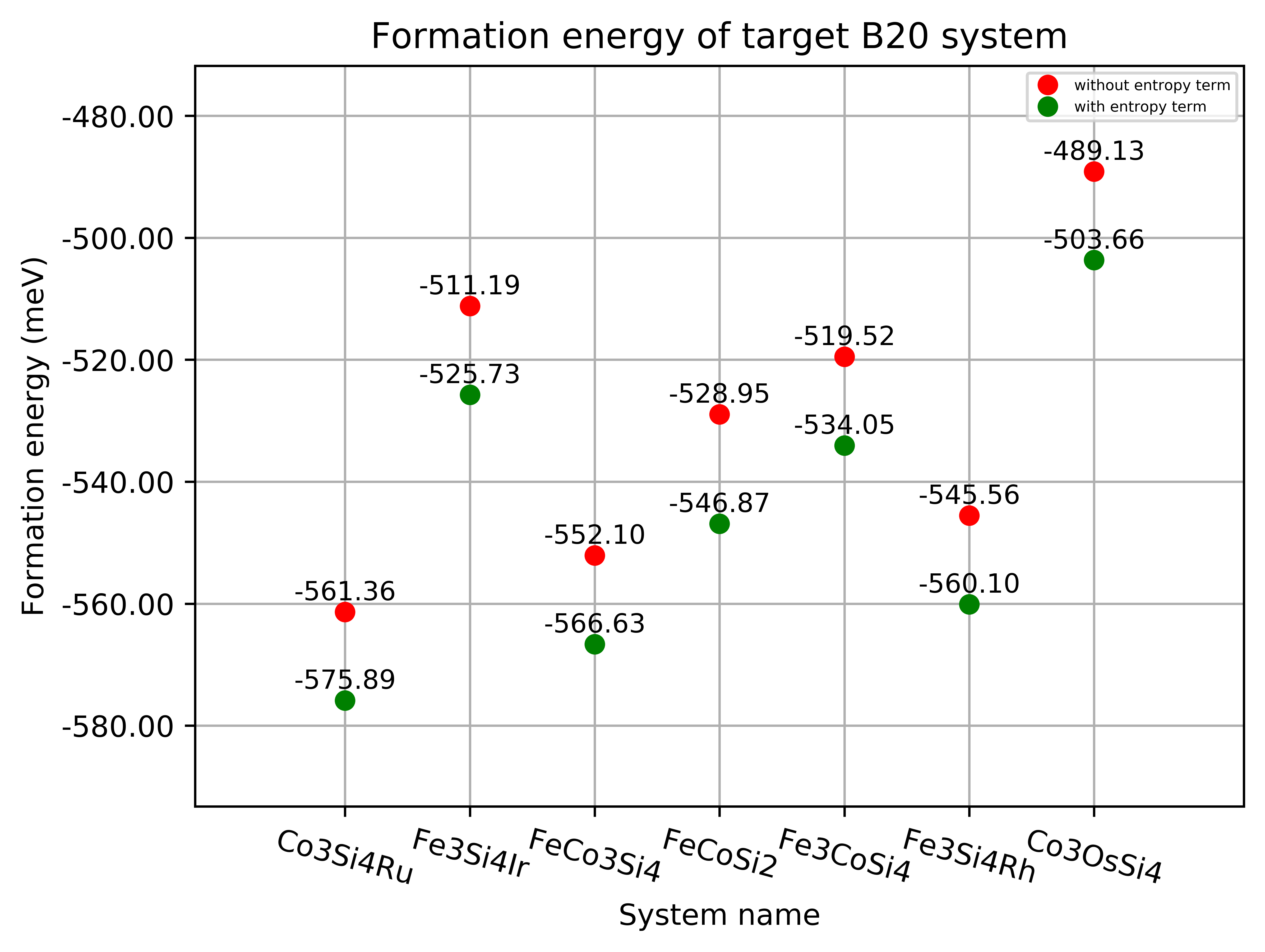}
\caption{Formation energy with and without mixing entropy term}
\label{f:formation_energy}
\end{figure}

The structural and phase stability of the doped B20 compounds was analyzed following the procedure depicted schematically in Figure~\ref{f:convex_hull_workflow}. The resulting convex-hull diagrams for all studied systems are shown in Figure~\ref{f:convex_hull_diagrams} where, for the sake of clarity, only the competing phases closest to the hull are indicated together with their distance from the hull. The formation energies of the doped B20 compounds are summarized in Figure~\ref{f:formation_energy}, where the values without and with the entropy contribution are provided.

The Quantum Espresso calculations of the formation energy (Figure~\ref{f:convex_hull_workflow}) were based on spin-polarized density functional theory (DFT) and SSSP efficiency pseudopotential library (version 1.1) \cite{Lejaeghereaad3000}, which shows good performance for the property prediction of magnetic materials. The Brillouin zone integration was done on a $(9\times 9\times 9)$ $k$-mesh, while the cutoff energy and the convergence threshold were set to $\unit[70]{Ry}$ and $\unit[10^{-6}]{Ry}$. The exchange-correlation energy was included within the Perdew-Burke-Ernzerhof (PBE) approximation. Spin-orbit coupling was not included in these calculations.

\newpage
\section{Trends in $\mathrm{Fe}_{1-x}\mathrm{Co}_{x}\mathrm{Si}$ and $\mathrm{Co}_{1-x}\mathrm{Ru}_{x}\mathrm{Si}$ series}
\label{SI_magnetic}

Here, we compare the magnetic trends for the Fe$_{1-x}$Co$_x$Si and Co$_{1-x}$Ru$_x$Si compounds, based on our first-principles results for $x=\frac14,\frac12,\frac34$, and discuss what can be expected from the future magnetic measurements for the Ru-doped compounds. Table~\ref{tab:FeCoSi_CoRuSi_trends} shows the concentration-dependence of different magnetic properties, such as the total magnetic moment (per formula unit), Curie temperature and $\nicefrac{D}{A}$ ratio for these two compound series.

As already mentioned in the introduction, the measured moment of Fe$_{1-x}$Co$_x$Si is rather small and reaches a maximum of $\unit[0.2]{\mu_\mathrm{B}}$ at $x=40\%$ \cite{Onose2005}. This is in contrast to our first-principles results, which suggest a monotonous decrease of moment for higher Co content between 25\% and 75\%. Also the maximal calculated value of $\unit[0.7]{\mu_\mathrm{B}}$ is much larger than the measured value, which leads to the overestimation of the Curie temperature by a factor of 4-5. This discrepancy can be attributed to the general drawback of DFT calculations which do not include the effect of spin fluctuations. As discussed in the main text, this drawback should not affect the conclusions on the $\nicefrac{D}{A}$ ratio, since both $A$ and $D$ scale as the square of the magnetic moment.

\begin{figure}
\centering
\includegraphics[width=0.99\columnwidth]{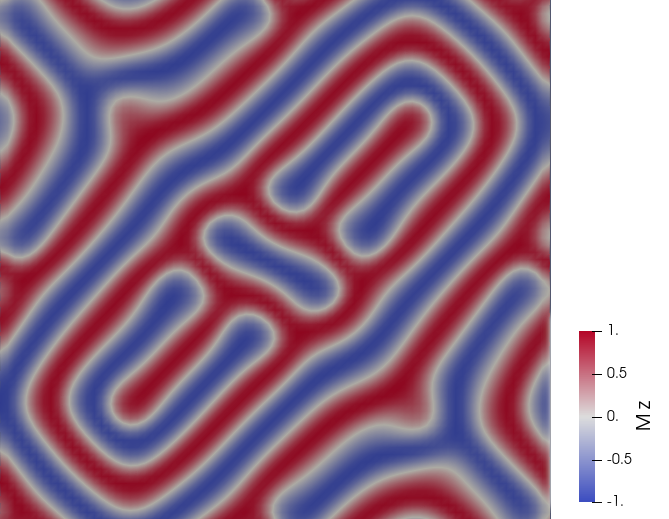}
\caption{The zero-field magnetic ground state of Fe$_{0.50}$Co$_{0.50}$Si calculated with the multiscale module $\mu$-\textsc{ASD}. The color palette indicates the $z$-component of the magnetic moment, where the red (blue) color represents the moments pointing fully along the +$z$-direction (-$z$-direction). The size of the simulation box is $(3\times3\times1)\,\mu$m.}
\label{fig:Fe2Co2Si4_texture}
\end{figure}

In contrast, the Co$_{1-x}$Ru$_{x}$Si series shows non-monotonic variation of total moment as a function of Ru concentration, where the moment at $x=50\%$ is almost twice the values at $x=25\%$ and $x=75\%$. This behavior is similar to the experimental observations for Fe$_{1-x}$Co$_{x}$Si \cite{Onose2005}, although the magnitude of the total moment is still probably overestimated. Nevertheless, based on the similarity between the Fe$_{1-x}$Co$_{x}$Si and Co$_{1-x}$Ru$_{x}$Si systems, one may expect this non-monotonic behavior of the magnetization vs Ru concentration to be actually observed in future measurements. Previous experimental studies~\cite{Grigoriev2013} revealed a sign change of the skyrmion chirality in Mn$_{1-x}$Fe$_x$Ge at $x\approx 75\%$ and this observation was attributed to the sign change of the DM interaction. Interestingly, our calculations show a similar effect for the spiralization $D$ and symmetric anisotropic exchange $\Gamma^{\alpha\beta}$ for the Fe$_{1-x}$Co$_x$Si and the new Co$_{1-x}$Ru$_{x}$Si series at moderate dopant concentrations. This is another important conclusion of our theoretical study that can be verified by future experimental work. Concerning the zero-field magnetic textures, all studied Fe$_{1-x}$Co$_x$Si and Co$_{1-x}$Ru$_{x}$Si compounds ($\frac14 \leq x \leq \frac34$) show helical magnetic orders which are qualitatively similar (Figure~\ref{fig:magnetic_textures}a,d,e and~\ref{fig:Fe2Co2Si4_texture}).

\setlength{\tabcolsep}{7pt}

\begin{table*}
 \caption{Trends in the calculated exchange stiffness $A$ and spiralization $D$ defined by Equation~(\ref{e:micromagnetic_parameters}) for the doped B20 compounds Fe$_{1-x}$Co$_x$Si and Co$_{1-x}$Ru$_x$Si. In addition, the Curie temperature and the total magnetic moment at $T=0$ are reported. For Co$_{0.50}$Ru$_{0.50}$Si the Curie temperature could not be estimated, and the ground state was found to be antiferromagnetic.
 }
 \vspace{5pt}
 \centering
 \begin{tabular}{l|cccc|ccc}
   \hline
    Compound &       $A$        &         $D$     & $\Gamma$ & $|D/A|$   &  $T_c$ & $M$  \\
     & $\mathrm{meV}\cdot\mathrm{\AA}^2$ & $\mathrm{meV}\cdot\mathrm{\AA}$  & $\mathrm{meV}\cdot\mathrm{\AA}^2$ & \AA$^{-1}$  &   $K$  & $\mu_\mathrm{B}/\mathrm{f.u.}$ \\
    \hline
    Fe$_{0.75}$Co$_{0.25}$Si  &  214.4  &  $-0.43$   &  +0.16  & 0.0020 & 240 & 0.702 \\
    Fe$_{0.50}$Co$_{0.50}$Si  &  159.9  &  $-0.51$  &  +0.11  & 0.0032 & 200 & 0.513 \\
    Co$_{0.75}$Fe$_{0.25}$Si  &   90.4  &  +0.38   & $-0.12$ & 0.0042 &  52 & 0.259 \\
    \hline
    Co$_{0.75}$Ru$_{0.25}$Si  &  81.2  &  $-0.56$ & $-0.29$  & 0.0069 & 56  & 0.252 \\
    Co$_{0.50}$Ru$_{0.50}$Si  &  51.6  &  +0.57   &  $-0.52$ & 0.0110 &    & 0.491 \\
    Co$_{0.25}$Ru$_{0.75}$Si  &  73.7  &  +1.00   &  +0.74   & 0.0136 &  48  & 0.250 \\
 \end{tabular}
 \label{tab:FeCoSi_CoRuSi_trends}
\end{table*}

\section{Phase analysis and crystal structure}

\begin{figure}
    \centering
    \includegraphics[width=0.99\columnwidth]{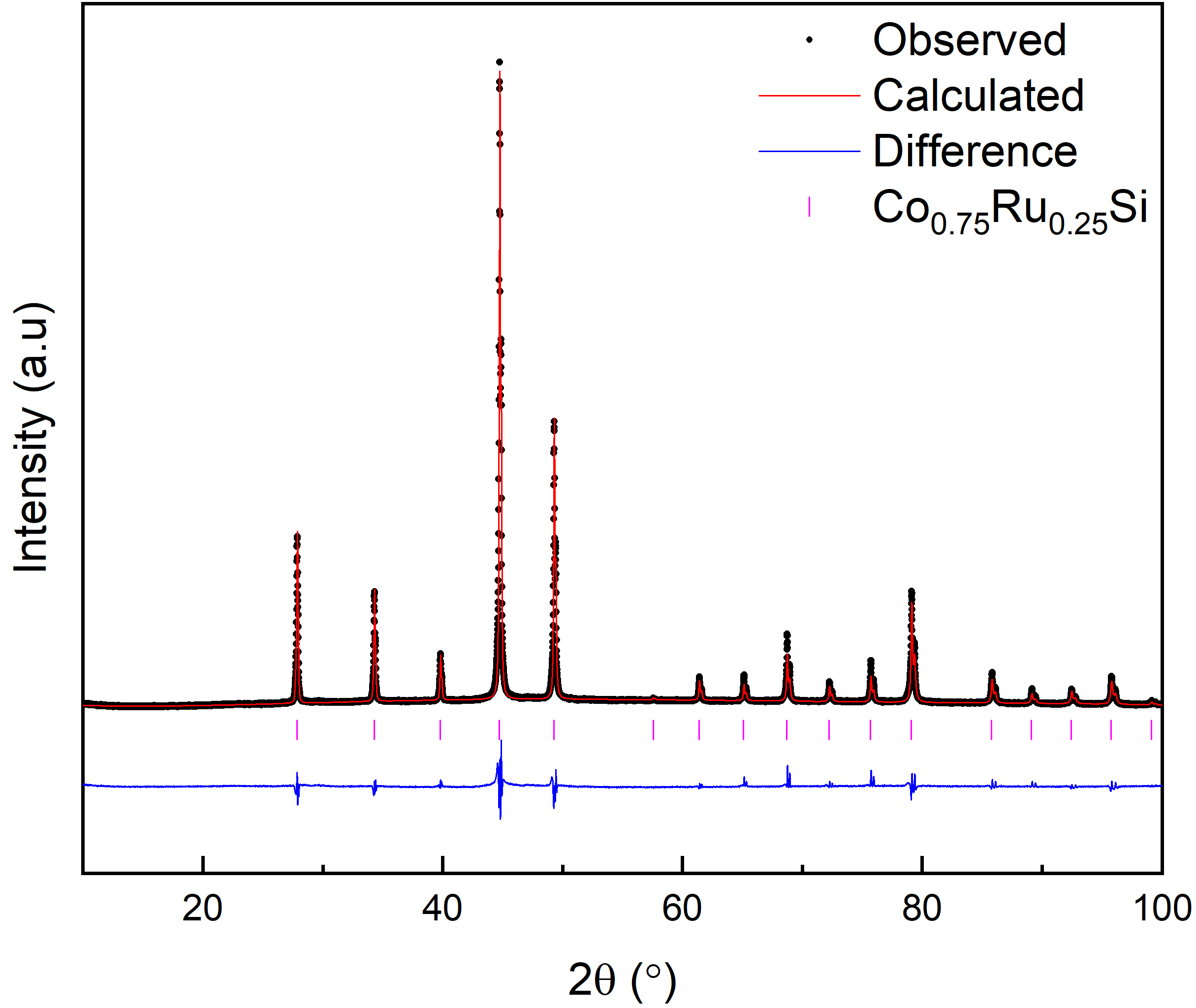}
    \vspace{-10pt}
    \caption{Powder X-ray diffraction pattern and Rietveld refinement of Co\textsubscript{0.75}Ru\textsubscript{0.25}Si annealed at $\unit[1773]{K}$. R\textsubscript{p} = 2.78, R\textsubscript{wp} = 4.28, $\chi$\textsuperscript{2} = 4.34. }
    \label{fig:Co0.75Ru0.25Si_Rietveld_Plot}
\end{figure}

\begin{figure}
    \centering
    \includegraphics[width=0.99\columnwidth]{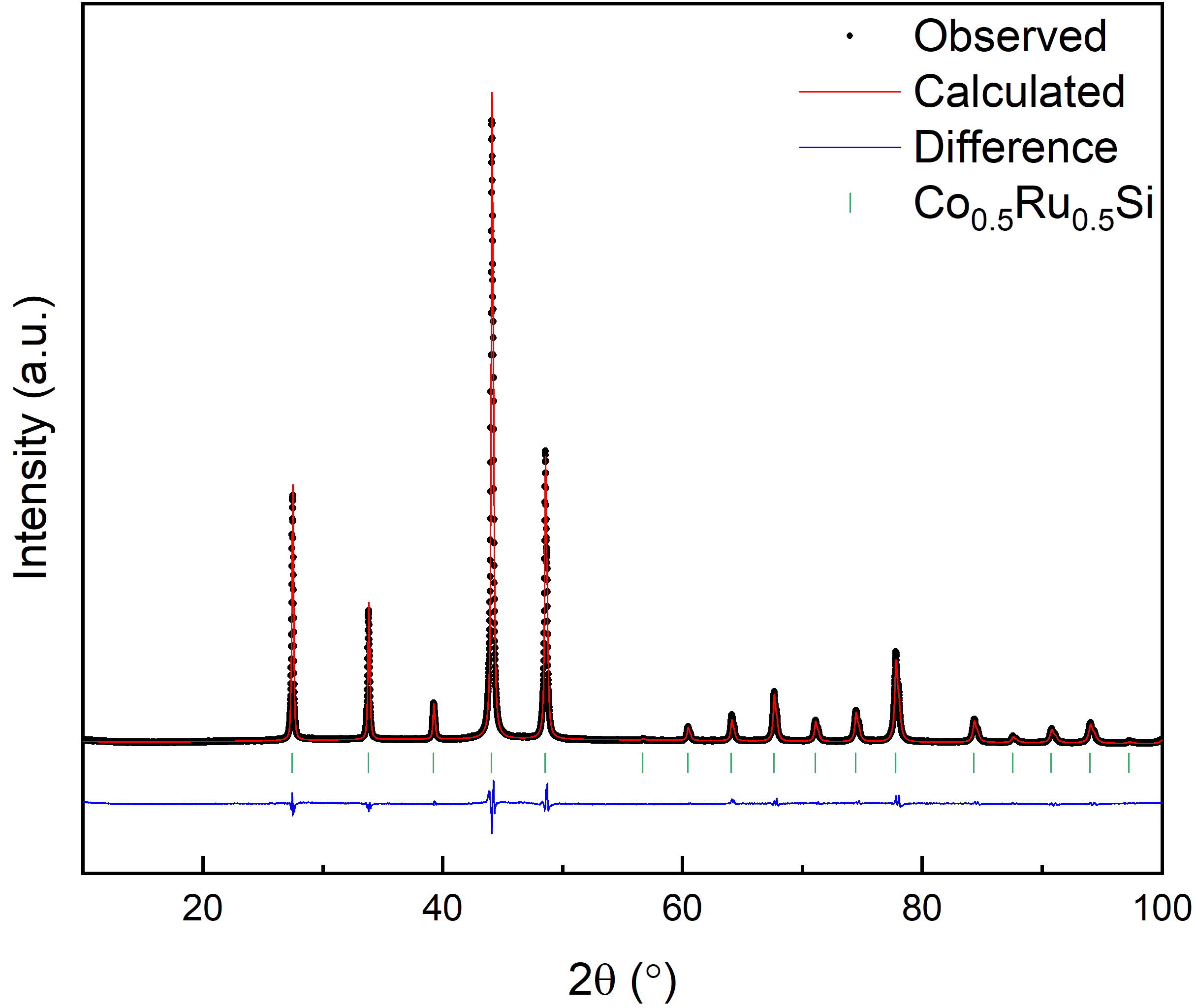}
    \caption{Powder X-ray diffraction pattern and Rietveld refinement of Co\textsubscript{0.5}Ru\textsubscript{0.5}Si annealed at $\unit[1773]{K}$. R\textsubscript{p} = 2.78, R\textsubscript{wp} = 3.56, $\chi$\textsuperscript{2} = 3.49. }
    \label{fig:Co0.5Ru0.5Si_Rietveld_Plot}
\end{figure}

\begin{figure}
    \centering
    \includegraphics[width=0.99\columnwidth]{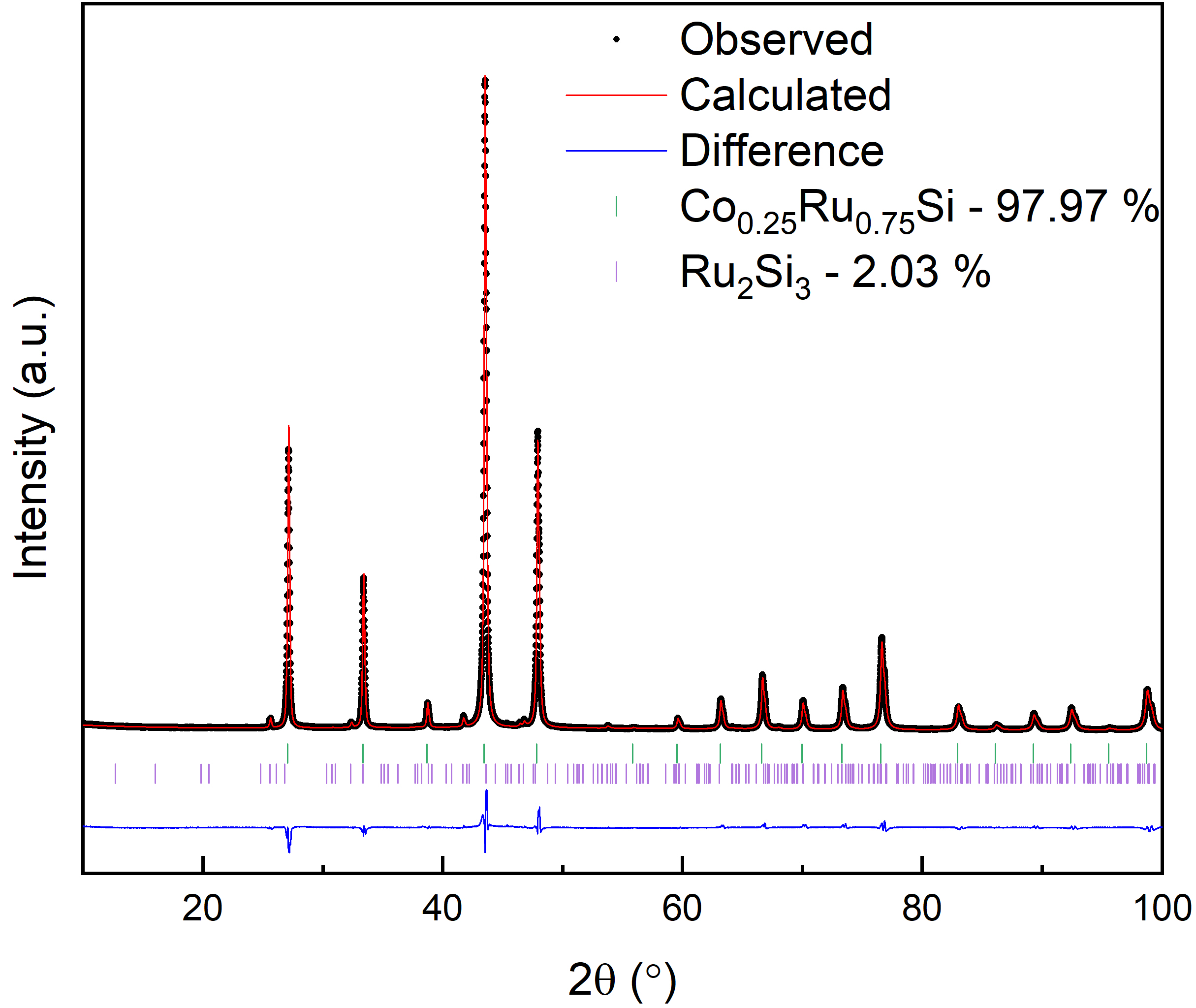}
    \caption{Powder X-ray diffraction pattern and Rietveld refinement of Co$_{0.25}$Ru$_{0.75}$Si annealed at $\unit[1773]{K}$. R\textsubscript{p} = 4.26, R\textsubscript{wp} = 5.53, $\chi$\textsuperscript{2} = 4.89. }
    \label{fig:Co0.25Ru0.75Si_Rietveld_Plot}
\end{figure}

\begin{figure}
    \centering
    \includegraphics[width=0.99\columnwidth]{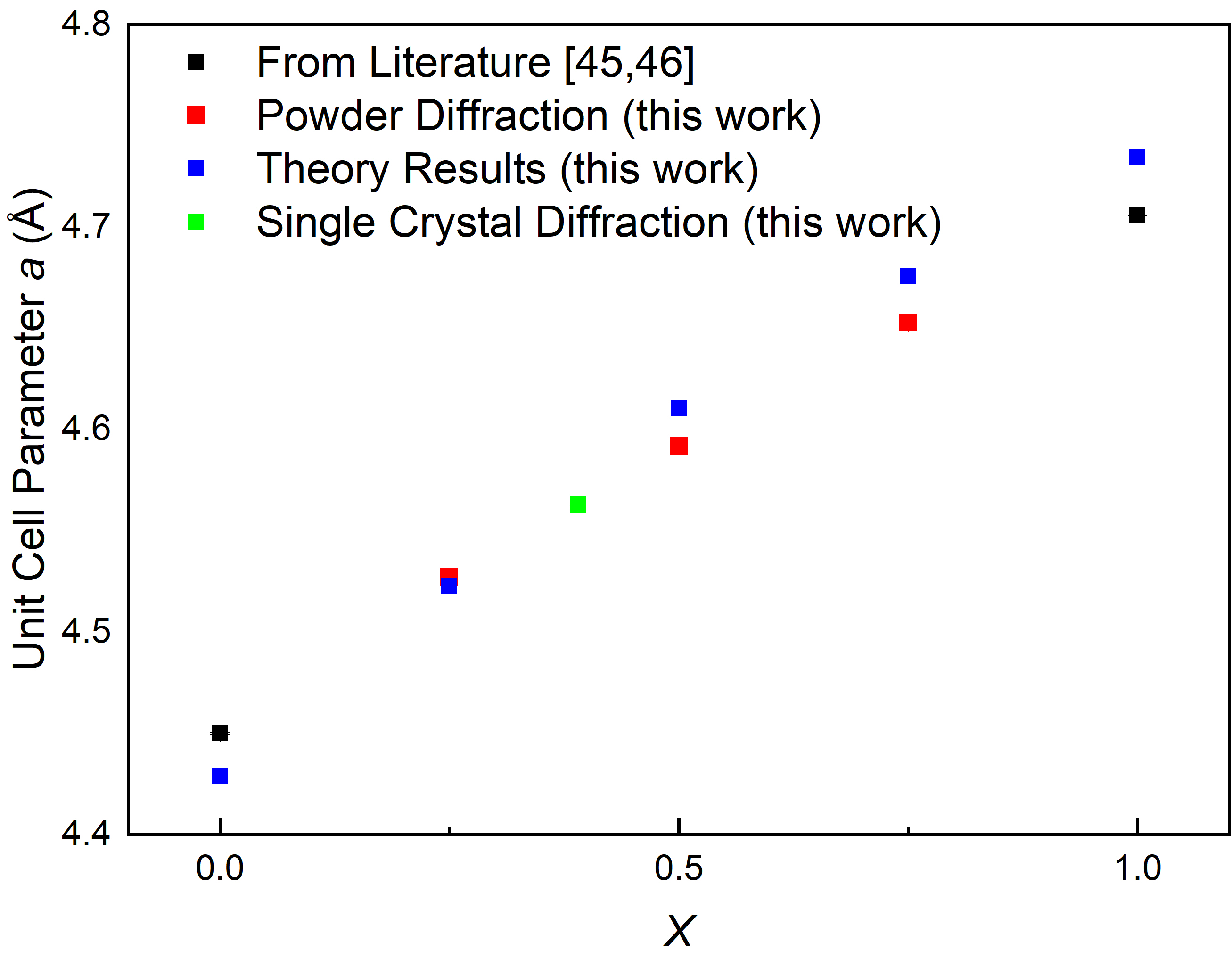}
    \caption{Variation of the unit cell parameter ($a$) with Ru concentration ($x$) in Co$_{1-x}$Ru$_x$Si derived from powder X-ray diffraction pattern data in the literature \cite{Weitzer1997,Demchenko2008}, powder diffraction (Table~\ref{tab:Atomic_Coordinates_Co0.75}--\ref{tab:Atomic_Coordinates_Co0.25}) and single-crystal diffraction (Table~\ref{tab:Co0.61Ru0.39Si_singlecrystal} and \ref{tab:C00.75Ru0.25Si Single crystal Coordinates}) and theoretical calculations in this work.}
    \label{fig:CoRuSi Unit Cells.jpg}
\end{figure}

The structural characterisation of the three new compositions, Co\textsubscript{0.75}Ru\textsubscript{0.25}Si, Co\textsubscript{0.5}Ru\textsubscript{0.5}Si and Co\textsubscript{0.25}Ru\textsubscript{0.75}Si is given in full in this section. The powder X-ray diffraction patterns refined with the Rietveld method are shown in Figure~\ref{fig:Co0.75Ru0.25Si_Rietveld_Plot}-\ref{fig:Co0.25Ru0.75Si_Rietveld_Plot}, the corresponding refined atomic positions and occupancies are given in Table~\ref{tab:Atomic_Coordinates_Co0.75}-\ref{tab:Atomic_Coordinates_Co0.25}. The refined unit cell parameters from X-ray diffraction data are plotted in Figure~\ref{fig:CoRuSi Unit Cells.jpg} alongside data from literature and theoretical values, obtained in this work using density functional theory, for comparison. Fairly good agreement between the measured and calculated lattice parameters for the whole series Co$_{1-x}$Ru$_x$Si shows the accuracy of the chosen theoretical approach. Co\textsubscript{0.75}Ru\textsubscript{0.25}Si was also analysed by single crystal X-ray diffraction, the refinement and structural details are shown in full in Table~\ref{tab:Co0.61Ru0.39Si_singlecrystal} and \ref{tab:C00.75Ru0.25Si Single crystal Coordinates}. A summary of the EDS data for each of the three new compositions is shown in Table~\ref{tab:EDS Data}, the table contains averaged compositions of data collected over at least 10 points per sample.

\setlength{\tabcolsep}{5pt}
\begin{table}
\caption{Atomic coordinates for Co\textsubscript{0.75}Ru\textsubscript{0.25}Si derived from Rietveld refinement of powder X-ray diffraction data.}
    \centering
    \begin{tabular}{lc|ccc|c}
        \hline
        Atom & site & $X$ & $Y$ & $Z$ & Occ\\
        \hline
        Co & 4a & 0.1387 (2) & 0.1387 (2) & 0.1387 (2) & 0.8\\
        Ru & 4a & 0.1387 (2) & 0.1387 (2) & 0.1387 (2) & 0.2 \\
        Si & 4a & 0.8403 (3) & 0.8403 (3) & 0.8403 (3) & 1.0\\
        \hline
    \end{tabular}
    \label{tab:Atomic_Coordinates_Co0.75}
\end{table}

\begin{table}
\caption{Atomic coordinates for Co\textsubscript{0.5}Ru\textsubscript{0.5}Si derived from Rietveld refinement of powder X-ray diffraction data}
    \centering
    \begin{tabular}{lc|ccc|c}
        \hline
        Atom & site & $X$ & $Y$ & $Z$ & Occ\\
        \hline
        Co & 4a & 0.1357 (1) & 0.1357 (1) & 0.1357 (1) & 0.6 \\
        Ru & 4a & 0.1357 (1) & 0.1357 (1) & 0.1357 (1) & 0.4 \\
        Si & 4a & 0.8396 (2) & 0.8396 (2) & 0.8396 (2) & 1.0\\
        \hline
    \end{tabular}
    \label{tab:Atomic_Coordinates_Co0.5}
\end{table}

\begin{table}
\caption{Atomic coordinates for Co\textsubscript{0.25}Ru\textsubscript{0.75}Si derived from  Rietveld refinement of powder X-ray diffraction data.}
    \centering
    \begin{tabular}{lc|ccc|c}
        \hline
        Atom & site & $X$ & $Y$ & $Z$ & Occ\\
        \hline
        Co & 4a & 0.1345 (1) & 0.1345 (1) & 0.1345 (1) & 0.3\\
        Ru & 4a & 0.1345 (1) & 0.1345 (1) & 0.1345 (1) & 0.7\\
        Si & 4a & 0.8419 (2) & 0.8419 (2) & 0.8419 (2) & 1.0 \\
        \hline
    \end{tabular}
    \label{tab:Atomic_Coordinates_Co0.25}
\end{table}

\begin{table}
    \caption{Crystallographic data and refinement details derived from single crystal X-ray diffraction data for Co$_{0.61}$Ru$_{0.39}$Si.}
    \centering
    \begin{tabular}{lc}
    \hline
    Formula & Co$_{2.44}$Ru$_{1.56}$Si$_{4}$\\
    \hline
    Formula weight & 413.2 \\
    Density (g/cm\textsuperscript{3}) & 7.223 \\
    Crystal system & cubic \\
    $a$ (Å) & 4.5629 (6) \\
    Volume (Å\textsuperscript{3}) & 95.00 (4) \\
    Z & 1 \\
    Measured reflections & 305 \\
    Independent reflections & 65 \\
    Refined parameters & 9 \\
    GOOF & 1.12 \\
    Final R indices (I$>$2$\sigma$(I)) & R1 = 0.0218, wR2 = 0.0465 \\
    \hline
    \end{tabular}
    \label{tab:Co0.61Ru0.39Si_singlecrystal}
\end{table}

\begin{table}
\caption{Atomic coordinates, site occupancies and equivalent isotropic displacement parameters (\AA$\textsuperscript{2}\times10\textsuperscript{3}$) of Co\textsubscript{0.75}Ru\textsubscript{0.25}Si derived from single crystal X-ray diffraction data.}
    \centering
    \begin{tabular}{lc|ccc|cc}
        \hline
         Atom & Site & $X$ & $Y$ & $Z$ & Occ & U\textsubscript{eq}  \\
         \hline
         Co & 4a & 0.3623 (2) & 0.3623 (2) & 0.3623 (2) & 0.61 & 2 (1)\\
         Ru & 4a & 0.3623 (2) & 0.3623 (2) & 0.3623 (2) & 0.39 & 2 (1)\\
         Si & 4a & 0.6582 (4) & 0.6582 (4) & 0.6582 (4) & 1.00 & 1 (2)\\
         \hline
    \end{tabular}
    \label{tab:C00.75Ru0.25Si Single crystal Coordinates}
\end{table}

\begin{table}
 \caption{EDS analysis of Co$_{1-x}$Ru$_x$Si, values given in At\%.}
    \centering
    \begin{tabular}{l|ccc}
        \hline
        Nominal Composition & Co & Ru & Si  \\
         \hline
         Co\textsubscript{0.75}Ru\textsubscript{0.25}Si & 38.17 (0.17) & 12.69 (0.18) & 49.14 (0.16)\\
         Co\textsubscript{0.50}Ru\textsubscript{0.50}Si & 25.13 (0.35) & 25.04 (0.40) & 49.83 (0.16) \\
         Co\textsubscript{0.25}Ru\textsubscript{0.75}Si & 13.68 (0.28) & 36.84 (0.35) & 49.49 (0.20) \\
         \hline
    \end{tabular}
    \label{tab:EDS Data}
\end{table}

\end{document}